\documentclass[fleqn,usenatbib]{mnras}

\usepackage{newtxtext,newtxmath}
\usepackage{graphicx} 
\usepackage{graphicx}	
\usepackage{amsmath}	
\usepackage{url}
\usepackage{subcaption}
\usepackage{booktabs}
\usepackage[normalem]{ulem}
\useunder{\uline}{\ul}{}
\usepackage{multirow}
\usepackage{comment}
\usepackage{pdflscape}

\usepackage[T1]{fontenc}

\DeclareRobustCommand{\VAN}[3]{#2}
\let\VANthebibliography\thebibliography
\def\thebibliography{\DeclareRobustCommand{\VAN}[3]{##3}\VANthebibliography}

\usepackage{graphicx}	
\usepackage{amsmath}	
\usepackage[acronym]{glossaries}

\newacronym{gw}{GW}{gravitational wave}
\newacronym{bh}{BH}{black hole}
\newacronym{wd}{WD}{white dwarf}
\newacronym{ns}{NS}{neutron star}
\newacronym{ms}{MS}{main sequence}
\newacronym{gs}{GS}{giant star}
\newacronym{hg}{HG}{Hertzsprung gap}
\newacronym{bbh}{BBH}{binary black hole}

\newacronym{pn}{PN}{post-Newtonian}

\newacronym{sn}{SN}{supernova}
\newacronym{ce}{CE}{common envelope}
\newacronym{rlo}{RLO}{Roche lobe overflow}
\newacronym{ppi}{PPI}{pulsational pair instability}
\glsdisablehyper

\newcommand{\kms}{\mathrm{km~s^{-1}}}
\newcommand{\density}{\mathrm{M_{\sun}~pc^{-3}}}

\title[BH star binaries in clusters]{Properties of black hole-star binaries formed in $N$-body simulations of massive star clusters:  implications for Gaia black holes}

\author[F. Fantoccoli et al.]{
Federico Fantoccoli,$^{1;2}$\thanks{f.fantoccoli@campus.unimib.it}
Jordan Barber,$^{1}$\thanks{barberj2@cardiff.ac.uk}
Fani Dosopoulou,$^{1}$
Debatri Chattopadhyay,$^{1}$
Fabio Antonini$^{1}$
\\
$^{1}$Gravity Exploration Institute, School of Physics and Astronomy, Cardiff University, Cardiff CF24 3AA, UK\\
$^{2}$Departement of Physics G. Occhialini, University of Milano Bicocca, Piazza della Scienza 3, I-20126 Milan, Italy,\\
}

\date{Accepted XXX. Received YYY; in original form ZZZ}

\pubyear{\the\year{}}

\begin{document}
\label{firstpage}
\pagerange{\pageref{firstpage}--\pageref{lastpage}}
\maketitle

\begin{abstract}
We investigate black hole-star binaries formed in $N$-body simulations of massive, dense star clusters. We simulate 32 clusters with varying initial masses ($10^{4}~\mathrm{M_{\odot}}$ to $10^{6}~\mathrm{M_{\odot}}$), densities ($1200~\mathrm{M_{\odot}pc^{-3}}$ to $10^{5}~\mathrm{M_{\odot}~pc^{-3}}$), and metallicities ($Z=0.01, 0.001, 0.0001$). Our results reveal that star clusters produce a diverse range of BH-star binaries, with dynamical interactions leading to extreme systems characterised by large orbital separations and high black hole masses. Of the ejected BH-main sequence (BH-MS) binaries, $20\%$ form dynamically, while the rest originate from the primordial binary population initially present in the cluster. Ejected BH-MS binaries that are dynamically formed  have more massive black holes, lower-mass stellar companions, and  over half are in a hierarchical triple system. 
All unbound BH-giant star (BH-GS) binaries were ejected as BH-MS binaries and evolved into the BH-GS phase outside the cluster. Due to their lower-mass companions, most dynamically formed binaries do not evolve into BH-GS systems within a Hubble time. Consequently, only 2 of the 35 ejected BH-GS binaries are dynamically formed. 
We  explore the formation pathways of Gaia-like systems, identifying two Gaia BH1-like binaries that formed through dynamical interactions, and two  Gaia BH2-like systems with a primordial origin. We did not find any system resembling Gaia BH3, which may however be attributed to the limited sample size of our simulations. 
\end{abstract}

\begin{keywords}
stars: kinematics and dynamics, methods: numerical, stars: black holes, globular clusters: general, galaxies: star clusters: general
\end{keywords}



\section{Introduction}

Black hole  binaries, particularly 
black hole-star (BH-S) systems, have long been essential for studying the formation and evolution of massive stars \citep[e.g.,][]{2006epbm.book.....E, 2006csxs.book..157M}. Historically, most of our understanding of \glspl{bh} comes from the study of X-ray binaries, where the \gls{bh} mass could be dynamically measured (e.g.,\cite{Ozel_2010, Farr_2011}). These systems represent an important observational benchmark for theoretical models of \gls{bh} formation. In 2015, the LIGO–Virgo collaboration detected the merger of two \glspl{bh}, marking the start of a new era in \gls{bh} research \citep{abbott_observation_2016}. This breakthrough has since led to the detection of more than 100 gravitational-wave (GW) events from \gls{bh} and \gls{ns} mergers \citep{Abbott_2021_a, Abbott_2021_b}, greatly expanding our understanding of \gls{bh} populations. 

In addition to \gls{gw} discoveries, recent observations have revealed a new family of quiescent, or X-ray silent, \glspl{bh} in binary systems. These quiescent \glspl{bh}, discovered through astrometry and radial velocity measurements  \citep[e.g.,][]{2019Sci...366..637T,2019Natur.575..618L}, have opened new windows into the study of non-interacting \gls{bh} binaries. The Gaia satellite, in particular, has been instrumental in identifying these systems. Recent Gaia discoveries include three quiescent \gls{bh} binaries: Gaia BH1 \citep{el2023sun, chakrabarti2023noninteracting}, Gaia BH2 \citep{el2023red, tanikawa2023search}, and Gaia BH3 \citep{gaia_collaboration_discovery_2024}. These systems, with their unique orbital characteristics and associations with distinct Galactic populations, challenge traditional models of \gls{bh} formation, particularly those involving isolated stellar binary evolution. 

The observed properties of Gaia \glspl{bh} are reported in Table~\ref{tab:Resume_Gaia}.
Gaia BH1 is a binary with a Sun-like star orbiting a \gls{bh} with a mass $\sim 10~\rm M_\odot$, an eccentricity $e \simeq 0.45$, and a period of $186~\rm days$. Gaia BH2 is a binary with a red giant orbiting a $9~\rm M_{\sun}$ \gls{bh}, with an eccentricity $e\simeq 0.5$, and a relatively long period of $1280~\rm days$.
Studies have investigated the formation of Gaia BH1 and Gaia BH2 through isolated mechanisms \citep{kotko2024enigmatic,kruckow2024formation,gilkis2024gaia}, however their peculiar orbital properties, such as long orbital periods and high eccentricity might be a challenge for these binary evolution models. For this reason, several previous studies using $N$-body simulations have investigated the possibility that these types of systems were dynamically formed in open star clusters \citep{shikauchi2020gaia,rastello2023dynamicalformationgaiabh1,Di_Carlo_2024,Tanikawa_CBCForm}.

Gaia BH3 is composed of a $33~\rm M_{\sun}$ \gls{bh} with a metal-poor giant star of mass $0.76~\rm M_{\sun}$ and metallicity [Fe/H]$\simeq -2.6$. The binary has an eccentricity of $e\simeq0.73$, and a long period $\simeq 4250~\rm days$.
It has been shown that Gaia BH3 can be either explained as an outcome of the evolution of an isolated binary with low metallicity \citep{iorio2024boringhistorygaiabh3, El_Badry_2024}, or that it might have formed in a globular cluster through dynamical interactions \citep{Mar_n_Pina_2024}. Its low metallicity and chemical composition further suggest it may be part of the ED-2 stream, a remnant of a dissolved globular cluster \citep{Balbinot_2024}. {These discoveries highlight the significance of understanding the formation pathways of black hole-star (BH-S) systems, particularly in diverse environments such as the Galactic plane, stellar halo, and dense star clusters. They also emphasise the need for more comprehensive models that incorporate both stellar evolution and dynamical processes in the formation of black hole binaries.}

In this work we use detailed $N$-body simulations of dense and massive star clusters to characterise the properties of the BH-S binaries formed in these models. By using the highly efficient $N-$body code $\tt PeTar$ developed by \cite{wang_petar_2020}, we  extend  the explored parameter space to clusters with higher masses and densities than previous work \citep{rastello2023dynamicalformationgaiabh1, Di_Carlo_2024, Tanikawa_CBCForm}, including stellar evolution and a realistic initial population of massive binary stars. {We investigate whether dynamical interactions in these models can generate binaries with properties resembling those of the observed Gaia black holes, and we explore potential constraints on a cluster-origin hypothesis for these systems.}

The paper is organised as follows: in Section \ref{sec: simulations} we describe the simulations and their initial conditions. in Sections \ref{sec:BH-MS}, \ref{sec:BH-GS}, \ref{sec:BH-WD} and \ref{sec:NS-S} we describe the properties of the ejected sample of BH-\gls{ms}, BH-\gls{gs}, BH-\gls{wd} and \gls{ns}-S (NS-MS, NS-GS) binaries produced  in our simulations. In addition, we give a qualitative description of the influence of \gls{ce}, tidal interactions and triple dynamics on the orbital properties of these systems. 

In  Section~\ref{sec:App_GaiaBH}, we describe the formation of Gaia BH-like systems. In Section~\ref{sec: Discussion} we study the formation efficiency of BH-MS and BH-GS systems in our simulations and compare our results with  previous literature.

\begin{table*}
\centering 
\begin{tabular}{cccccccc} \hline
&   Stellar type
       & $M_{\rm BH} [\rm M_{\sun}]$  & $M_{*} [\rm M_{\sun}]$ &  $a [\rm R_{\sun}]$ &  $P [\rm days]$ & $e$ &  [Fe/H]  \\
\midrule
Gaia BH1 &  G-type main-sequence & $9.62_{-0.18}^{+0.18}$  & $0.93_{-0.05}^{+0.05}$ & $301.55_{-2.15}^{+2.15}$  & $185.59_{-0.05}^{+0.05}$ & $0.45_{-0.005}^{+0.005}$ & -0.2 \\[0.1cm]
Gaia BH2  & Red giant &$8.94_{-0.0.34}^{+0.34}$ & $1.07_{-0.19}^{+0.19}$ & $1066.55_{-17.20}^{+17.20}$ & $1276.7_{-0.6}^{+0.6}$ & $0.5176_{-0.0009}^{+0.0009}$  &-0.2 \\[0.1cm]
Gaia BH3 & G-type giant&$32.7_{-0.82}^{+ 0.82}$  & $0.76_{-0.05}^{+0.05}$& $2477.035_{58.058}^{+58.058}$ & $4253.1_{-98.5}^{+98.5}$ & $0.7291_{-0.0048}^{+0.0048}$ &-2.6 \\[0.1cm]
\bottomrule
\end{tabular}
\caption{\label{tab:Resume_Gaia} Main properties of  Gaia BH1 \citep{el2023sun}, Gaia BH2 \citep{el2023red}, and Gaia BH3 \citep{El_Badry_2024}.}
\end{table*}

\section{Methods}
\label{sec: simulations}
In this work we use the high-performance hybrid \textit{N}-body code $\tt PeTar$ \citep{wang_petar_2020} to simulate 32 stellar clusters with initial cluster masses $(M_{\rm c})$ ranging from $10^{4}~\mathrm{M_{\sun}}$ to $10^{6}~\mathrm{M_{\rm \sun}}$, initial half-mass densities $(\rho_{\rm h})$ from $1200~\density$ to $10^{5}~\density$ and metallicities $Z \in \{10^{-2},10^{-3},10^{-4}\}$. $\tt PeTar$  makes use of the $PT + PP$ method introduced in \citet{oshino_particleparticle_2011}, which combines a direct \textit{N}-body (particle-particle) method with a Barnes-Hut tree (particle-tree) method \citep{barnes_hierarchical_1986}. Few body interactions are computed using a combination of the fourth-order Hermite integrator and the slow-down time-transformed symplectic integrator \citep[$\tt SDAR$][]{wang_slow-down_2020}. In addition, $\tt PeTar$ exploits parallisation via a hybrid parallel method based on the $\tt FDPS$ framework \citep{iwasawa_implementation_2016, iwasawa_accelerated_2020, namekata_fortran_2018}. The simulation code is configured to make use of OpenMP and MPI processes for parallelisation and performs long-range force calculations with Nvidea P100 GPUs, all of which  reduce the computational cost compared to other $N$-body codes. {The lower computational cost enables us to more extensively explore the initial cluster parameter space, specifically allowing for the modelling of massive and dense clusters over multiple relaxation timescales.}

Stellar and binary evolution is included in our models and is computed using the binary stellar evolution packages $\tt SSE$ and $\tt BSE$ \citep{hurley_comprehensive_2000, hurley_evolution_2002, banerjee_bse_2020} . While $\tt PeTar$ does not directly compute any \gls{pn} terms within the equations of motion, compact objects can coalesce through \gls{gw} emission following the semi-major axis and eccentricity evolution as described in \citet{peters_gravitational_1964}.

\subsection{Initial conditions}
We initialise our cluster conditions using $\tt McLuster$ \citep{wang_complete_2019, kupper_mass_2011}, adopting a King density profile \citep{king_structure_1966} with a concentration parameter $W_{0}=8$. We do not include the effect of a  galactic tidal field in any of our simulations. Most models have an initial half-mass density $\rho_{\rm h}=1.2\times10^{3}~\rm \density$ as this is a typical value found for globular clusters in the Galaxy \citep{harris_catalog_2013}.
We also explore higher densities, $\rho_{\rm h}=10^{4}~\rm \density$ and $\rho_{\rm h}=10^{5}~\rm \density$, and vary the initial cluster mass  from $10^{4}~\rm M_{\sun}$ to $10^{6}~\rm M_{\sun}$. We consider three
values of metallicity: $Z=0.01,~0.001~{\rm and}~0.0001$.

For most cluster models we consider two variations. In one there are no binaries at the initialisation of the cluster and for the other variation we ensure every star with mass $\geq20~\mathrm{M_{\sun}}$ is initialised in a binary. {The latter choice is motivated by observations of young clusters and associations in which there is  a large binary fraction ($>70\%$) amongst these massive stars \citep{sana_binary_2012}. Binaries formed with the formation of the cluster are commonly termed \textit{primordial} binaries. We note that since our primordial binaries are always set such that the primary mass is $\geq20~\rm M_{\sun}$, we are not considering the formation of BH-S binaries through exchanges that involve a primordial binary in which both components are low-mass stars.} This is unlikely to affect significantly our results since such binaries are unlikely to undergo frequent dynamical interactions leading to an exchange of one of the components with a massive star or a \gls{bh}. Due to their lower mass these binaries do not efficiently migrate to the cluster core where these interactions can take place \citep{spitzer_dynamical_1987}. However, we note that we might be underestimating the number of BH-S binaries formed. 

To initialise the stellar masses of the cluster particles we adopt a \citet{kroupa_variation_2001} initial mass function with $0.08~{\rm M_{\sun}}\leq M \leq 150~{\rm M_{\sun}}$. Primordial binaries are then generated by taking every star with mass $\geq20~\mathrm{M_{\sun}}$ and finding an ideal companion mass by drawing from a uniform mass ratio ($q$) distribution with $0.1\leq q \leq1$; the particle in the cluster with the closest mass to what was drawn is then chosen as the binary partner.
The adopted uniform mass ratio distribution is consistent with what derived observationally \citep{moe_mind_2017}.  The eccentricity for these binaries is then drawn from a \citet{sana_binary_2012} distribution
\begin{equation}
    f_{\mathrm{e}}=0.55e^{-0.45}.
    \label{eq:eccdistinit}
\end{equation}
The binary period is set using the extended \citet{sana_binary_2012} distribution described in \citet{oh_dependency_2015}
\begin{equation}
    f_{\log_{10}(P)} = 0.23 \left[\log_{10}\left(\frac{P}{\rm days}\right)\right]^{-0.55}.
    \label{eq:periodinit}
\end{equation}
We draw \gls{sn} natal kicks from a Maxwellian distribution with $\sigma=265~\kms$ \citep{hobbs_statistical_2005} and assume a fallback kick prescription when scaling the kicks for \gls{bh} formation \citep{fryer_mass_1999}. In addition, we assume a \citet{fryer_compact_2012} rapid \gls{sn} engine for the compact object remnant mass, and strong \gls{ppi} cut-off at $45~\mathrm{M_{\sun}}$ \citep{belczynski_evolutionary_2020}. { Unlike previous studies \citep{rastello2023dynamicalformationgaiabh1,Di_Carlo_2024,El-badry2022}, we allowed  the binary to survive without  merging if a phase of common-envelope evolution  is initiated by a donor star that is on the Hertzsprung gap.}

The largest integration time simulated was $3~\rm Gyr$ with the majority of simulations integrated for $1~\rm Gyr$. The final integration time of the simulations is chosen such that it is several times the initial relaxation time of the cluster. For the most massive clusters ($M_{\rm c}=5\times 10^5~\mathrm{M_{\sun}}$ and $10^6~\mathrm{M_{\sun}}$), we make sure that the simulation runs for at least half the initial relaxation time. 

The first five columns of Table~\ref{tab:models} are adapted from \cite{barber2024} and provide a breakdown of the cluster initial conditions for each of our simulations.

\begin{table*}   
    \centering
    \resizebox{\textwidth}{!}{
  \begin{tabular}{ccccccccccc}
        &  &  &  &  & \multicolumn{3}{c}{Primordial} & \multicolumn{3}{c}{Dynamical} \\
        Model & Metallicity & Total Mass & Density & Binary Fraction & \multicolumn{3}{c}{BH-MS(BH-GS){[}BH-WD{]}NS-S} & \multicolumn{3}{c}{BH-MS(BH-GS){[}BH-WD{]}NS-S} \\
         &  & M$_{\sun}$ & $\mathrm{M_{\sun}~pc^{-3}}$ &  & Ejected & Retained & Total & Ejected & Retained & Total \\ \hline
        \multirow{2}{*}{Z1-M1-D3} & \multirow{8}{*}{0.01} & \multirow{2}{*}{10,000} & \multirow{2}{*}{1200} & 0 & 0(0){[}0{]}0 & 0(0){[}0{]}0 & 0(0){[}0{]}0 & 0(0){[}0{]}0 & 0(0){[}0{]}0 & 0(0){[}0{]}0 \\
         &  &  &  & 0.0025 & 1(1){[}0{]}0 & 2(2){[}0{]}0 & 3(3){[}0{]}0 & 0(0){[}0{]}0 & 1(1){[}0{]}0 & 1(1){[}0{]}0 \\
        \multirow{2}{*}{Z1-M5-D3} &  & \multirow{2}{*}{50,000} & \multirow{2}{*}{1200} & 0 & 0(0){[}0{]}0 & 0(0){[}0{]}0 & 0(0){[}0{]}0 & 1(0){[}0{]}0 & 0(0){[}0{]}0 & 1(0){[}0{]}0 \\
         &  &  &  & 0.0025 & 3(2){[}1{]}4 & 7(3){[}0{]}2 & 10(5){[}1{]}6 & 0(0){[}0{]}0 & 0(1){[}0{]}0 & 0(1){[}0{]}0 \\
        \multirow{2}{*}{Z1-M10-D3} &  & \multirow{2}{*}{100,000} & \multirow{2}{*}{1200} & 0 & 0(0){[}0{]}0 & 0(0){[}0{]}0 & 0(0){[}0{]}0 & 0(0){[}0{]}0 & 0(0){[}0{]}0 & 0(0){[}0{]}0 \\
         &  &  &  & 0.0026 & 3(2){[}1{]}5 & 29(25){[}0{]}2 & 32(27){[}1{]}7 & 1(0){[}1{]}0 & 2(0){[}0{]}0 & 3(0){[}1{]}0 \\
        Z1-M50-D3 &  & 500,000 & 1200 & 0 & 0(0){[}0{]}0 & 0(0){[}0{]}0 & 0(0){[}0{]}0 & 0(0){[}0{]}0 & 0(0){[}0{]}0 & 0(0){[}0{]}0 \\
        Z1-M100-D3 &  & 1,000,000 & 1200 & 0 & 0(0){[}0{]}0 & 0(0){[}0{]}0 & 0(0){[}0{]}0 & 0(0){[}0{]}0 & 0(0){[}0{]}0 & 0(0){[}0{]}0 \\ \hline
        \multirow{2}{*}{Z2-M1-D3} & \multirow{16}{*}{0.001} & \multirow{2}{*}{10,000} & \multirow{2}{*}{1200} & 0 & 0(0){[}0{]}0 & 0(0){[}0{]}0 & 0(0){[}0{]}0 & 0(0){[}0{]}0 & 0(2){[}0{]}0 & 0(2){[}0{]}0 \\
         &  &  &  & 0.0025 & 2(1){[}0{]}1 & 3(3){[}0{]}0 & 5(4){[}0{]}1 & 0(0){[}1{]}0 & 1(0){[}0{]}0 & 1(0){[}1{]}0 \\
        \multirow{2}{*}{Z2-M5-D3} &  & \multirow{2}{*}{50,000} & \multirow{2}{*}{1200} & 0 & 0(0){[}0{]}0 & 0(0){[}0{]}0 & 0(0){[}0{]}0 & 0(0){[}0{]}0 & 0(0){[}0{]}0 & 0(0){[}0{]}0 \\
         &  &  &  & 0.0025 & 2(0){[}1{]}0 & 17(25){[}0{]}0 & 19(25){[}1{]}0 & 0(0){[}0{]}0 & 0(0){[}0{]}0 & 0(0){[}0{]}0 \\
        \multirow{2}{*}{Z2-M5-D3-L} &  & \multirow{2}{*}{50,000} & \multirow{2}{*}{1200} & 0 & 0(0){[}0{]}0 & 0(0){[}0{]}0 & 0(0){[}0{]}0 & 0(0){[}0{]}0 & 0(0){[}0{]}0 & 0(0){[}0{]}0 \\
         &  &  &  & 0.0025 & 0(0){[}2{]}0 & 21(24){[}0{]}1 & 21(24){[}2{]}1 & 0(0){[}0{]}0 & 0(2){[}0{]}0 & 0(2){[}0{]}0 \\
        \multirow{2}{*}{Z2-M10-D3} &  & \multirow{2}{*}{100,000} & \multirow{2}{*}{1200} & 0 & 0(0){[}0{]}0 & 0(0){[}0{]}0 & 0(0){[}0{]}0 & 1(0){[}0{]}0 & 1(1){[}0{]}0 & 2(1){[}0{]}0 \\
         &  &  &  & 0.0026 & 9(4){[}1{]}9 & 47(49){[}0{]}2 & 56(53){[}1{]}11 & 0(0){[}0{]}0 & 3(2){[}0{]}0 & 3(2){[}0{]}0 \\
        \multirow{2}{*}{Z2-M10-D3-L} &  & \multirow{2}{*}{100,000} & \multirow{2}{*}{1200} & 0 & 0(0){[}0{]}0 & 0(0){[}0{]}0 & 0(0){[}0{]}0 & 3(0){[}0{]}0 & 0(0){[}0{]}0 & 3(0){[}0{]}0 \\
         &  &  &  & 0.0026 & 4(0){[}1{]}3 & 48(46){[}0{]}7 & 52(46){[}1{]}10 & 0(0){[}0{]}0 & 1(1){[}0{]}0 & 1(1){[}0{]}0 \\
        \multirow{2}{*}{Z2-M10-D4} &  & \multirow{2}{*}{100,000} & \multirow{2}{*}{10,000} & 0 & 0(0){[}0{]}0 & 0(0){[}0{]}0 & 0(0){[}0{]}0 & 1(0){[}0{]}0 & 2(1){[}0{]}0 & 3(1){[}0{]}0 \\
         &  &  &  & 0.0025 & 7(5){[}1{]}4 & 32(41){[}0{]}1 & 39(46){[}1{]}5 & 0(0){[}0{]}0 & 2(2){[}0{]}0 & 2(2){[}1{]}0 \\
        \multirow{2}{*}{Z2-M1-D5} &  & \multirow{2}{*}{10,000} & \multirow{2}{*}{100,000} & 0 & 0(0){[}0{]}0 & 0(0){[}0{]}0 & 0(0){[}0{]}0 & 0(0){[}0{]}0 & 0(1){[}0{]}0 & 1(0){[}0{]}0 \\
         &  &  &  & 0.0025 & 2(2){[}1{]}3 & 1(1){[}0{]}0 & 3(3){[}1{]}3 & 4(2){[}1{]}0 & 1(2){[}1{]}0 & 5(4){[}2{]}0 \\
        \multirow{2}{*}{Z2-M5-D5} &  & \multirow{2}{*}{50,000} & \multirow{2}{*}{100,000} & 0 & 0(0){[}0{]}0 & 0(0){[}0{]}0 & 0(0){[}0{]}0 & 3(0){[}1{]}0 & 1(2){[}0{]}0 & 4(2){[}1{]}0 \\
         &  &  &  & 0.0025 & 14(8){[}6{]}2 & 5(12){[}0{]}0 & 19(20){[}6{]}2 & 3(0){[}0{]}0 & 0(1){[}0{]}0 & 3(1){[}0{]}0 \\ \hline
        \multirow{2}{*}{Z3-M1-D3} & \multirow{8}{*}{0.0001} & \multirow{2}{*}{10,000} & \multirow{2}{*}{1200} & 0 & 0(0){[}0{]}0 & 0(0){[}0{]}0 & 0(0){[}0{]}0 & 0(0){[}0{]}0 & 2(1){[}0{]}0 & 2(1){[}0{]}0 \\
         &  &  &  & 0.0025 & 1(1){[}0{]}0 & 7(7){[}0{]}0 & 8(8){[}0{]}0 & 0(0){[}0{]}0 & 1(2){[}0{]}0 & 1(2){[}0{]}0 \\
        \multirow{2}{*}{Z3-M5-D3} &  & \multirow{2}{*}{50,000} & \multirow{2}{*}{1200} & 0 & 0(0){[}0{]}0 & 0(0){[}0{]}0 & 0(0){[}0{]}0 & 0(0){[}0{]}0 & 1(1){[}0{]}0 & 1(1){[}0{]}0 \\
         &  &  &  & 0.0025 & 4(2){[}1{]}3 & 34(34){[}0{]}2 & 38(36){[}1{]}5 & 0(0){[}0{]}0 & 1(5){[}0{]}0 & 1(5){[}0{]}0 \\
        \multirow{2}{*}{Z3-M10-D3} &  & \multirow{2}{*}{100,000} & \multirow{2}{*}{1200} & 0 & 0(0){[}0{]}0 & 0(0){[}0{]}0 & 0(0){[}0{]}0 & 0(0){[}0{]}0 & 0(0){[}0{]}0 & 0(0){[}0{]}0 \\
         &  &  &  & 0.0025 & 13(5){[}1{]}6 & 58(66){[}0{]}0 & 71(71){[}1{]}6 & 1(0){[}0{]}0 & 1(2){[}0{]}0 & 2(2){[}0{]}0 \\
        Z3-M50-D3 &  & 500,000 & 1200 & 0 & 0(0){[}0{]}0 & 0(0){[}0{]}0 & 0(0){[}0{]}0 & 0(0){[}0{]}0 & 0(0){[}0{]}0 & 0(0){[}0{]}0 \\
        Z3-M100-D3 &  & 1,000,000 & 1200 & 0 & 0(0){[}0{]}0 & 0(0){[}0{]}0 & 0(0){[}0{]}0 & 0(0){[}0{]}0 & 0(0){[}0{]}0 & 0(0){[}0{]}0 \\ \hline
        \multicolumn{2}{c}{Total} & 4,580,000 &  &  & 66(33){[}17{]}40 & 311(338){[}0{]}17 & 377(371){[}17{]}57 & 19(2){[}4{]}0 & 21(30){[}1{]}0 & 40(32){[}5{]}0
    \end{tabular}}
     \caption{Initial cluster conditions for our $\tt PeTar$ $N$-body simulations. Each model is given a unique name based on its initial setup (metallicity, initial cluster mass and density). Models with a -L added  are run for $3~\rm Gyr$; Z1-M100-D3 and Z3-M50-D3  are terminated at $507~\rm Myr$, and Z3-M100-D1  and Z3-M100-D3 at
     $254~\rm Myr$. All the remaining models are evolved up to a maximum integration time of $1~\rm Gyr$.     
     Each model contains two variations, one which starts with no binaries, and one which sets an initial binary fraction of 100\% amongst massive stars (initial mass $\geq20~\rm M_{\sun}$). }
    \label{tab:models}
\end{table*}

\subsection{Classification and post-ejection evolution}
\label{sec: methods}
In this section, we present the two classification criteria for the binaries. Firstly, we classify the binaries as \textit{ejected} binaries and \textit{retained} binaries. In this paper we will mostly focus on the ejected population, which are binaries that are ejected from the cluster either by a \gls{sn} kick or through a dynamical interaction. We focus on the ejected binaries since they are more relevant to the halo/field population observed by Gaia. However, we also provide a less detailed description of the retained binaries. There are two conditions that must be fulfilled to classify a binary as ejected from the cluster: firstly, we impose $r_{\rm bin}>20r_{\rm h}$, where $r_{\rm bin}$ is the position of the centre of mass of the binary and $r_{\rm h}$ is the cluster half-mass radius at any given time. The second criterion is that the velocity of the centre of mass of the binary is larger than the escape velocity from the cluster. {If these conditions are both satisfied at a given evolutionary stage (e.g., BH-MS), then the binary is classified as ejected. For the remaining binaries we only include them in our retained binary population after imposing a condition that the binary survives inside the cluster until at least the next stage of stellar evolution, e.g., a \gls{ms} star becoming a \gls{gs}.} Thus, a retained binary is a binary that is formed as a BH-type binary inside the cluster, and it evolves to another type within the cluster.

The second classification is to divide the ejected binaries in primordial and dynamical. Primordial are the binaries initially present in the cluster, though they may experience dynamical processes that change their orbit. Dynamical binaries are those in which the components are paired through  gravitational encounters during the simulations.
We take the binaries at the time of ejection and subsequently evolve them in isolation for a Hubble time, using the $\tt COSMIC$ code \citep{Breivik_2020}. All the stellar evolution parameters in $\tt COSMIC$ are set to be exactly the same as in  $\tt PeTar$ so that the stellar evolution is consistent through the evolution. 
However, we note that the isolated evolution of an ejected binary might not be a good approximation in some cases. If the binary is ejected in a stable multiple system (triple, quadruple or more), the isolated binary evolution set-up will ignore the effect of the tertiary (or other objects in a multiple system) on the binary evolution 
\citep{2022MNRAS.516.1406S}.

The total number of primordial and dynamical binaries for each of our simulations is provided Table~\ref{tab:models}, each split into the ejected and retained populations and then further split into the stellar type of the binary components. We consider binaries in which one component is a \gls{bh} and the other component is either a \gls{ms} star, a \gls{wd}, or a \gls{gs}. We also briefly consider binaries in which one component is a \gls{ns} and the other is a star (either a \gls{ms} or a \gls{gs}). We discuss in detail the properties of these systems in the following sections.

\begin{figure*} 
\includegraphics[width=1.6\columnwidth]{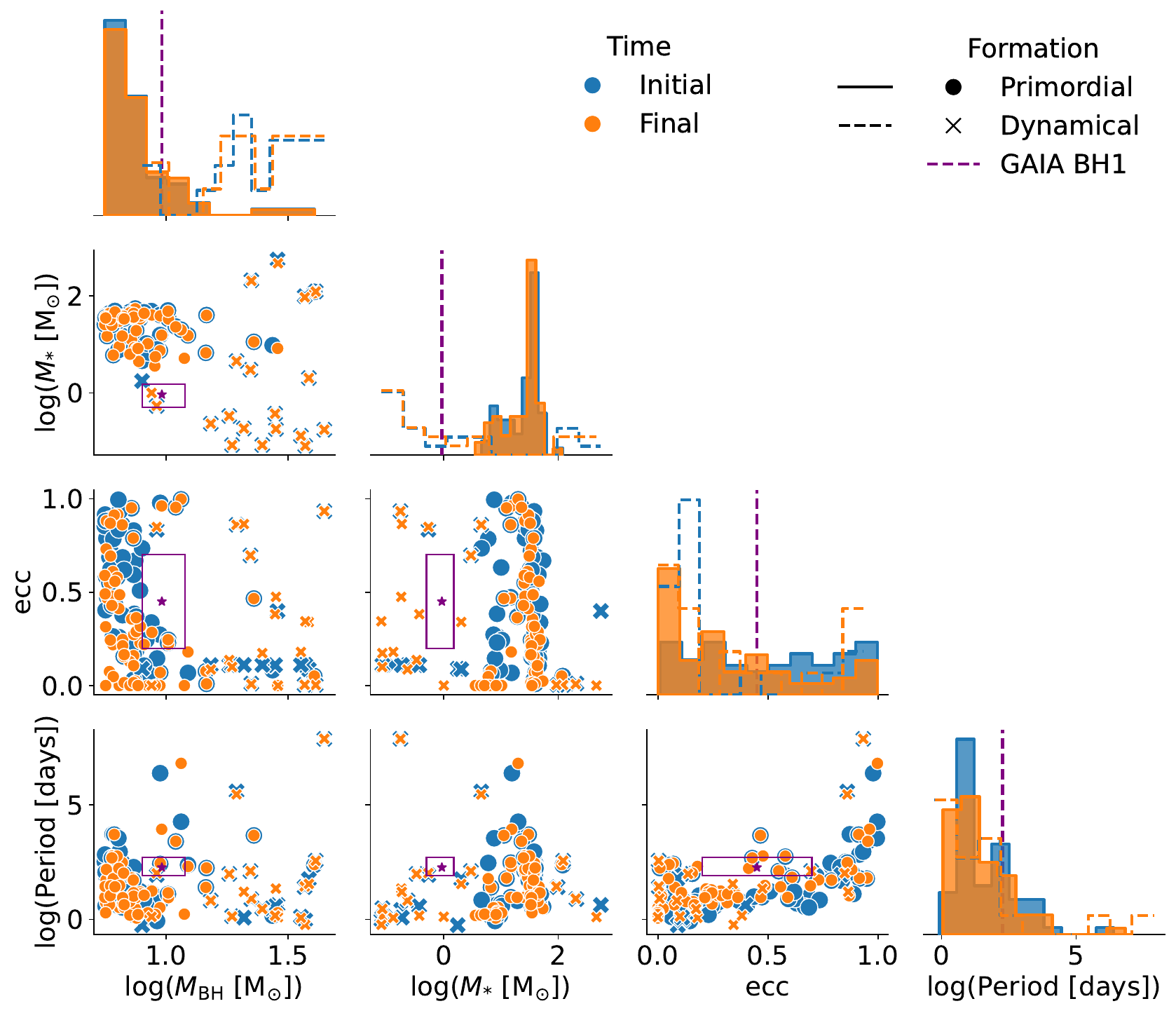}
\caption{Component mass and orbital properties of the ejected BH-MS binaries. In the scatter(histogram) plots, dots(lines) and crosses(dashes) represent primordial binaries and dynamical binaries, respectively; binaries at the first BH-MS time are in blue and at the last BH-MS time in orange. The purple contours define the Gaia BH1 similarity region. }
\label{f: BH-MS corner}
\end{figure*}

\section{BH-MS  binaries}
\label{sec:BH-MS}

In this section, we characterise the ejected BH-MS population; \gls{ms} stars are defined following the $\tt BSE$  classification of stellar types in \cite{hurley_evolution_2002}. The component mass and orbital properties of the ejected BH-MS binaries are shown in Fig.~\ref{f: BH-MS corner}.
We consider two evolutionary times when performing  the analysis: (i) the first time the stellar component is on the MS,  and (ii) the last time-step in which the stellar component  is classified as a  \gls{ms}. {The aim is to illustrate the range of parameter space that a system can explore throughout its lifetime as a BH-MS binary, considering that stellar evolution processes, tidal interactions, and mass transfer events may still influence the properties of the binary.} 
The contours represent the 'region of similarity' in the parameter space for the Gaia BH1 system. They are defined by the following limits: $M_{*} \in [0.5, 1.5] ~\rm M_{\sun}$, $M_{\rm BH} \in [8, 12] ~\rm M_{\sun}$, $ e \in [0.2, 0.7]$, $a \in [200, 600] ~\rm R_{\sun}$, $P \in [80, 500]~\rm days$. These contours are much wider than the errors on the Gaia BH1-system parameters (reported in Table~\ref{tab:Resume_Gaia}), typically smaller than $2\%$, and highlight the \emph{Gaia BH1-like binaries} -- for a similar analysis in lower mass cluster simulations see \cite{rastello2023dynamicalformationgaiabh1}.

The distributions of the \gls{bh} and \gls{ms} star masses for dynamical and primordial  binaries show that the two populations are well separated. For primordial binaries, the \gls{bh} mass distribution is peaked around $M_{\rm BH} \simeq 7 ~\rm M_{\sun}$, while for dynamically formed binaries the distribution is nearly uniform between $10 ~\rm M_{\sun}$ and $40 ~\rm M_{\sun}$. This discrepancy underlines the possibility for dynamically formed binaries to cover a wider range of mass ratios  compared to  primordial binaries. The \gls{ms} mass distribution for dynamical and primordial binaries is also distinctly separated. The primordial binary distribution has a strong peak around $M_{*} \simeq 30 ~\rm M_{\sun}$, while the dynamical population produces essentially no binaries near this peak, extending to both lower and higher masses and containing a large fraction of lower mass stars with a peak at  around $M_{*} \simeq 0.45 ~\rm M_{\sun}$.
We find that $93\%$  of the ejected BH-MS binaries are formed inside the cluster. 
The eccentricity and period distributions for primordial and dynamical binaries cover a similar range of values, with no differing features.
Moreover, we do not find significant difference in the distributions between the initial and final time of the \gls{ms} evolutionary stage.

Fig.~\ref{f: BH-MS corner} shows that the eccentricity distribution has a peak at $e \simeq 0$, and above 
$e \simeq 0$ the distribution appears nearly uniform. 
The peak could be due to a \gls{ce} phase; $18\%$ of BH-MS ejected system progenitors undergo a \gls{ce} phase before ejection.
As a caveat we note that post common envelope systems might have eccentricities up to $\sim0.2$ and that the evolution in this phase remains uncertain \citep{kruckow2021catalog}. In addition to a \gls{ce} phase,  tidal forces in the binary contribute to the circularisation of the orbit \citep{hurley_evolution_2002,Hut_tides}. The strength of both equilibrium and dynamical tides depends on the ratio between the radius of the star and the semi-major axis of the orbit ($R / a$), and the binary mass ratio. Given these scales, we can approximately track the impact of tides by searching for \gls{rlo} events in the binary sample. During a \gls{rlo} phase, the radius of one of the two stars (or both) strongly increases, enhancing the effects of tides and leading to orbital circularisation.
We find that $68\%$ of the ejected BH-MS binaries progenitors undergo at least one \gls{rlo} phase before  ejection. It is important to underline that all the BH-MS progenitors that undergo a \gls{ce} phase, previously start a \gls{rlo} overflow event. Therefore, the orbit of  $26\%$ of the systems that undergo a \gls{rlo} is immediately  circularised due to a \gls{ce} phase. These binaries  are the low eccentricity population shown in Fig.~\ref{f: BH-MS corner}.

{We note, however, that \cite{Sepinsky2007,Dosopoulou_2016_a, Dosopoulou_2016_b}, showed that tidal interactions do not always result in rapid circularisation during the early stages of mass transfer. As a result, mass transfer at periastron in eccentric orbits may introduce significant eccentricities.}

\begin{figure*} \includegraphics[width=2\columnwidth]{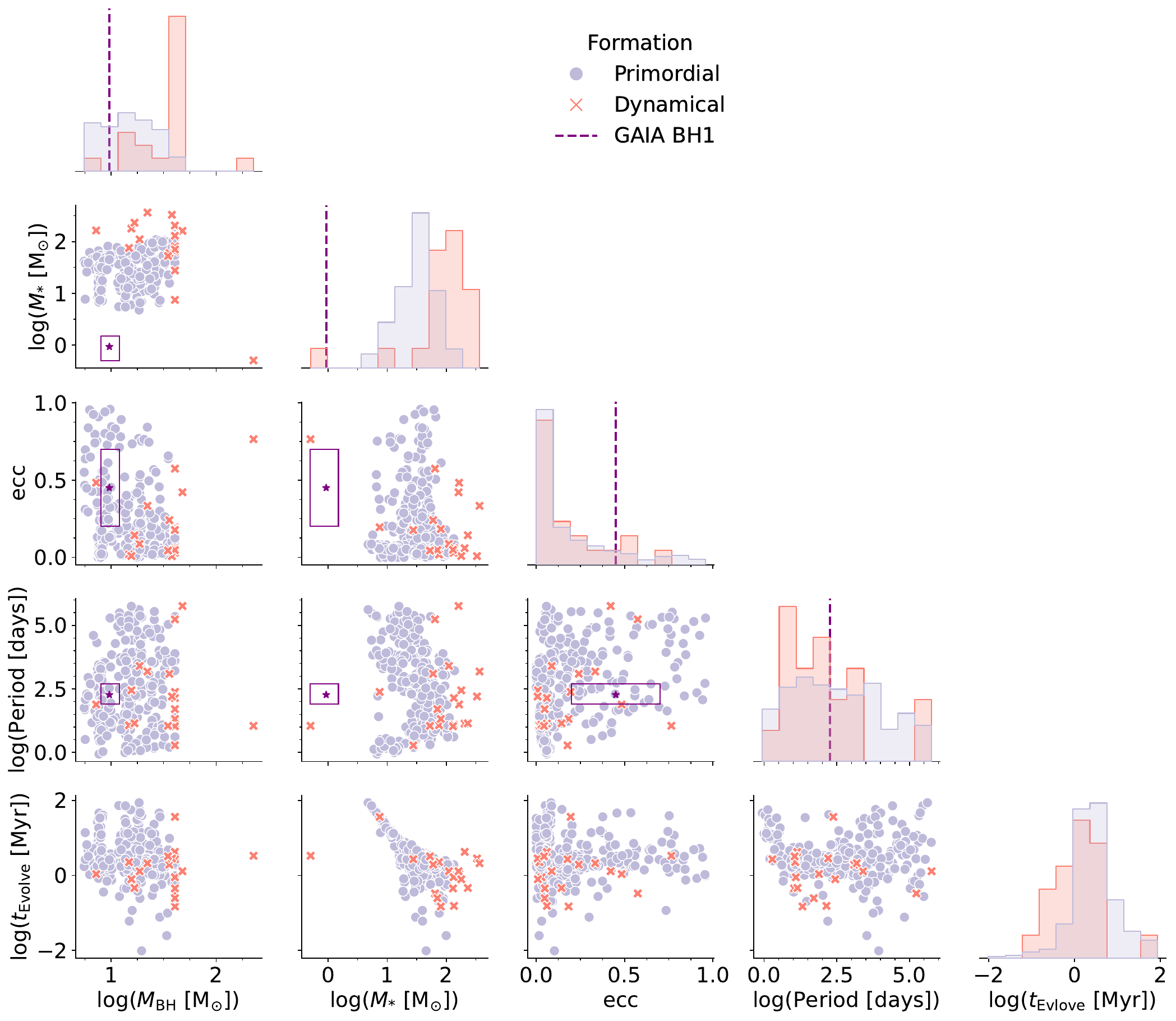} \caption{Properties of BH-MS retained binaries. Red dots and the blue crosses represent primordial binaries and dynamical binaries, respectively; the purple contours depict the Gaia BH1 similarity region. Here $t_{\rm evolve}$ represents the lifetime of the binary, i.e., the time between formation and when the stellar component evolves off the MS.} \label{f: BH-MS corner_RET} \end{figure*}

In addition, we underline that \gls{ce} evolution and tides are not the only processes that affect the eccentricity distribution: the eccentricity distribution still remains nearly uniform due to kicks during \gls{bh} formation  (though attenuated in our prescription) or due to dynamical interactions  in the cluster. 
Moreover, a more detailed analysis of the effect of tides {and mass transfer}  is needed to fully address their role in our models, but this is beyond the scope of this work.

After ejection, no BH-MS binary experiences a \gls{ce} phase (the \gls{ce} phase generally occurs as the star is leaving the \gls{ms}). However, $32\%$ of the ejected BH-MS undergo at least one \gls{rlo} event; this shifts more binaries towards lower eccentricities in the final population, explaining the stronger peak at $e\sim0$ in the orange distribution in Fig.~\ref{f: BH-MS corner}. 

As previously mentioned, the presence of a tertiary companion can alter the evolution of the inner BH-MS binary (or its progenitors), leading to a different evolutionary path. In addition, \citet{Tanikawa_CBCForm} has shown that Gaia BH-like binaries are often found with an accompanying tertiary star. Therefore we look for triple companions to the BH-MS binaries before the time of ejection. We find that $42\%$ of the binaries have been in a {stable} triple system before they are ejected from the cluster. We define a \textit{stable} triple as a system that has an outer eccentricity $e < 1$. we find that $37\%$ of the stable triples occur when the inner binary is a BH-MS. Moreover, in the majority of these systems ($85\%$), the inner binary is dynamically formed. When analysing the properties of the triples in our models, we  look at the last evolution snapshot in which  the system is present in the simulation: they have an average semi-major axis ratio $a_{\rm in}/a_{\rm out} = 0.015$ (where $a_{\rm in}$ is the semi-major axis of the inner orbit and $a_{\rm out}$ of the outer orbit), and the  average outer orbital eccentricity is $e = 0.93$.

Out of the 85 ejected BH-MS binaries, only 13 are found in stable triple systems. This implies that in at least 85$\%$ of cases, treating the binary as isolated is a reasonable approximation for its evolution after ejection from the cluster. {We find that in 11 of the 13 ejected triple systems the inner binary is from the dynamical population. Thus, $68\%$ of the ejected dynamical BH-MS binaries, and $3\%$ of the ejected primordial BH-MS binaries are in a stable triple at ejection.} This suggests that the presence of a tertiary companion is a signature of dynamical formation in our models.

\subsection{BH-MS retained binaries}
\label{sec:BH-MS_Ret}
We consider now  the properties of the retained BH-MS binaries in our simulations, the properties of which are shown in Fig.~\ref{f: BH-MS corner_RET}. We note that we found no substantial change between the main properties of BH-MS binaries at the time of formation and at the last BH-MS time. Therefore, we show only the binaries taken at the final BH-MS time. {However, we do keep the distinction between primordial and dynamical binaries and find that $94\%$ of the binaries have a primordial origin, while $6\%$ are formed dynamically.}

The distributions shown in Fig.~\ref{f: BH-MS corner_RET} are similar to those in Fig.~\ref{f: BH-MS corner}.
The \gls{bh} mass distribution shows an evident cut-off at $\simeq 45~\rm M_{\sun}$.  This sharp mass limit is due to \gls{ppi} \gls{sn} that prevents the formation of any \glspl{bh} above this mass value. 
The few binaries that cross the threshold have a dynamical origin, and are either formed through hierarchical \gls{bbh} mergers or through the accretion of a massive star by a \gls{bh} \citep{barber2024}. It is important to point out that currently $\tt PeTar$ does not compute a \gls{gw} recoil kick following a \gls{bbh} merger. Therefore, some of the massive \glspl{bh} formed through hierarchical mergers that we find, will likely be ejected from the cluster shortly after forming and thus not go on to form future binaries.

{We note that the population of dynamically formed binaries contains a significant sub-population with $M_{*}\lesssim 1~\rm M_{\sun}$ -- the median \gls{ms} mass of this clump is $\simeq 0.44~\rm M_{\sun}$. These binaries are formed early during the simulation and, in almost all cases, are immediately disrupted: the median formation time is $ \simeq 4.11~\rm Myr$ from the beginning of the simulation, and the median disruption time is the same. Therefore most of them undergo an immediate disruption after formation and are not shown in Fig. \ref{f: BH-MS corner_RET}.

On the other hand, for ejected BH-MS binaries, the population of the corresponding low mass clump is made of 13 binaries: 3 binaries survive as BH-MS for an average time $\simeq 971~\rm Myr$, the other 10 stay BH-MS until the end of the simulation. The most similar binary to Gaia BH1 (see Section~\ref{sec:Gaia_BH1 closest}) is part of this long-lived population. }

\begin{figure*}
\includegraphics[width=1.6\columnwidth]{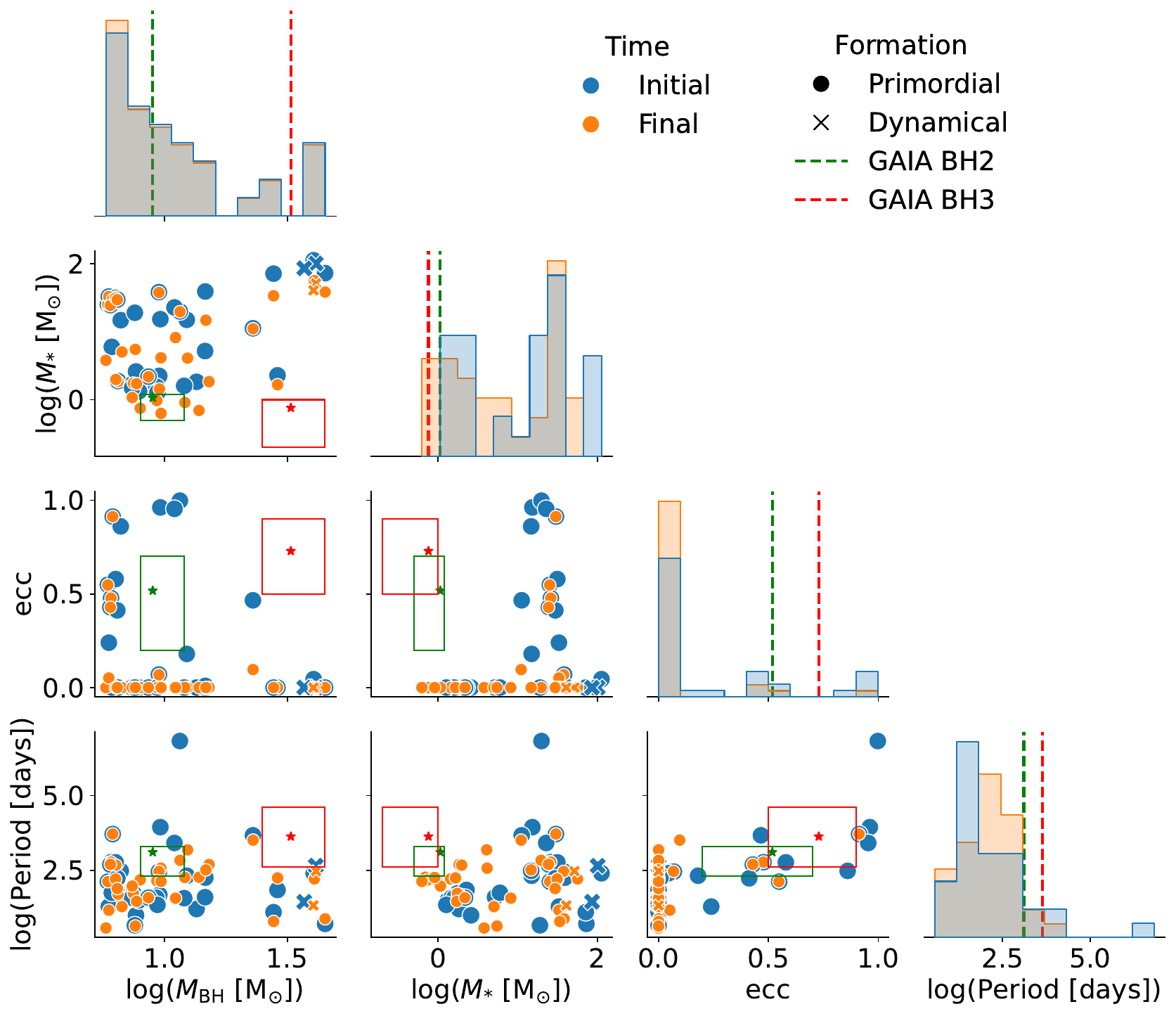}
\caption{Properties of the BH-GS  ejected binaries produced in our models. The dots and the crosses represent primordial binaries and dynamical binaries, respectively. The binaries at the first BH-GS time are shown in blue, the last BH-GS time is represented by orange colour. The green and red contours are the Gaia BH2 and Gaia BH3 similarity regions.}
\label{f: BH-GS corner}
\end{figure*}

\section{BH-GS  Binaries}
\label{sec:BH-GS}
In this section, we describe the BH-GS binary population in our models; \glspl{gs} are defined following the $\tt BSE$ classification as stars with  indices 3, 4, or 5 \citep[e.g.,][]{Di_Carlo_2024}. The results are reported in Fig.~\ref{f: BH-GS corner}. The Gaia BH2 similarity region is defined as: $M_{*} \in [0.5, 1.2]~\rm M_{\sun}$, $M_{BH} \in [8, 12]~\rm M_{\sun}$, $ e \in [0.2, 0.7]$, $a \in [238, 1088]~\rm R_{\sun}$, $P \in [200, 2000]~\rm days$, while for Gaia BH3 $M_{*} \in [0.2, 1]~\rm M_{\sun}$, $M_{\rm BH} \in [25, 45]~\rm M_{\sun}$, $ e \in [0.5, 0.9]$, $a \in [560, 11760]~\rm R_{\sun}$, $P \in [400, 40000]~\rm days$; the same contours for Gaia BH3 are adopted in \cite{Mar_n_Pina_2024}. All the ejected BH-GS binaries in our models are first ejected as BH-MS binaries and then evolve to a BH-GS binary outside the cluster.

\begin{figure*}
\includegraphics[width=2\columnwidth]{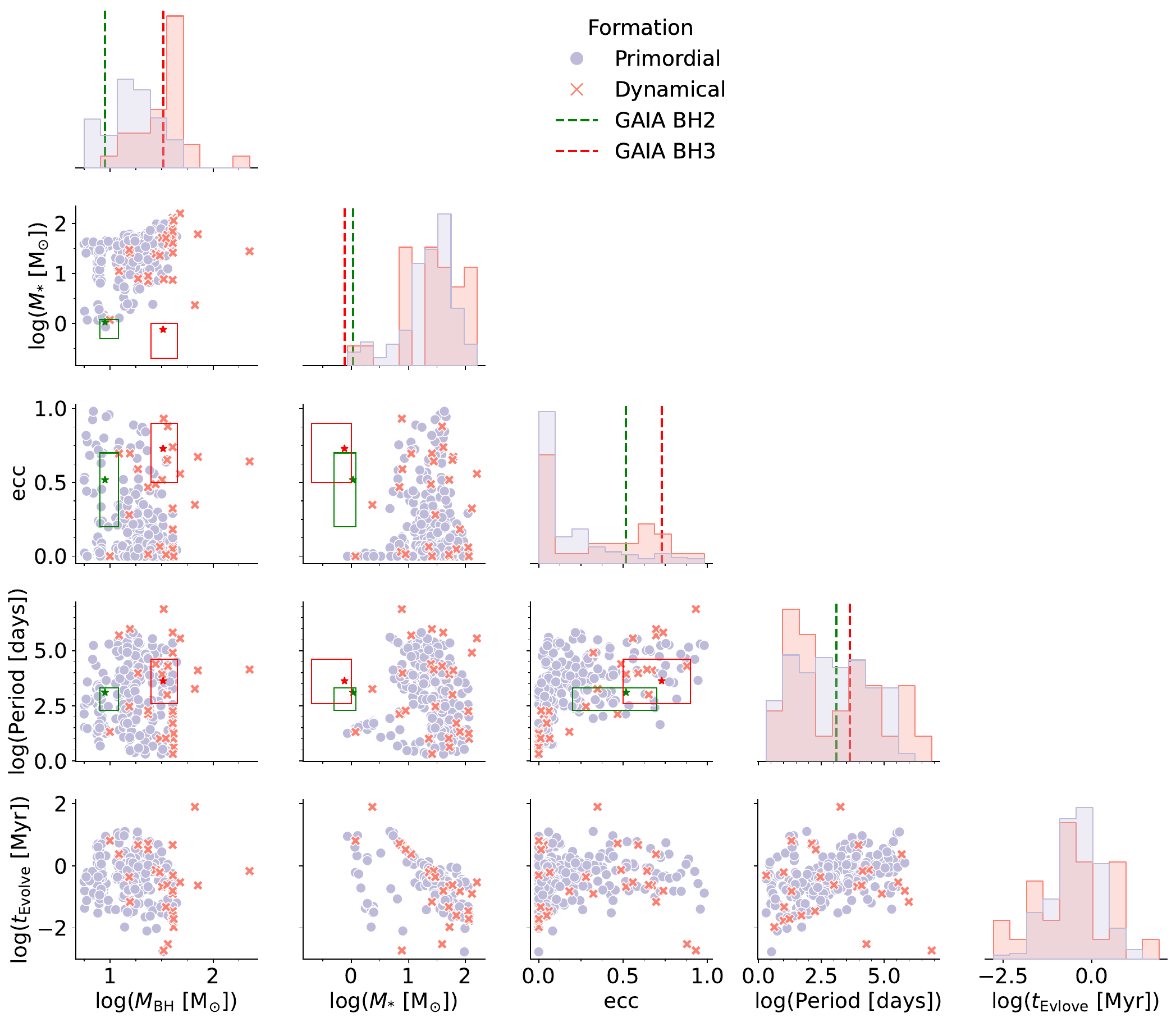}
\caption{ Properties of the retained BH-GS binaries. The red dots and the blue crosses represent primordial binaries and dynamical binaries respectively; the green and red contours are the Gaia BH2 and Gaia BH3  similarity region, respectively. Here $t_{\rm evolve}$ represents the lifetime of the binary, i.e., the time between formation and 
 when the stellar component evolves off the giant phase.}
\label{f: BH-GS corner_RET}
\end{figure*}

\begin{figure*}
\includegraphics[width=1.6\columnwidth]{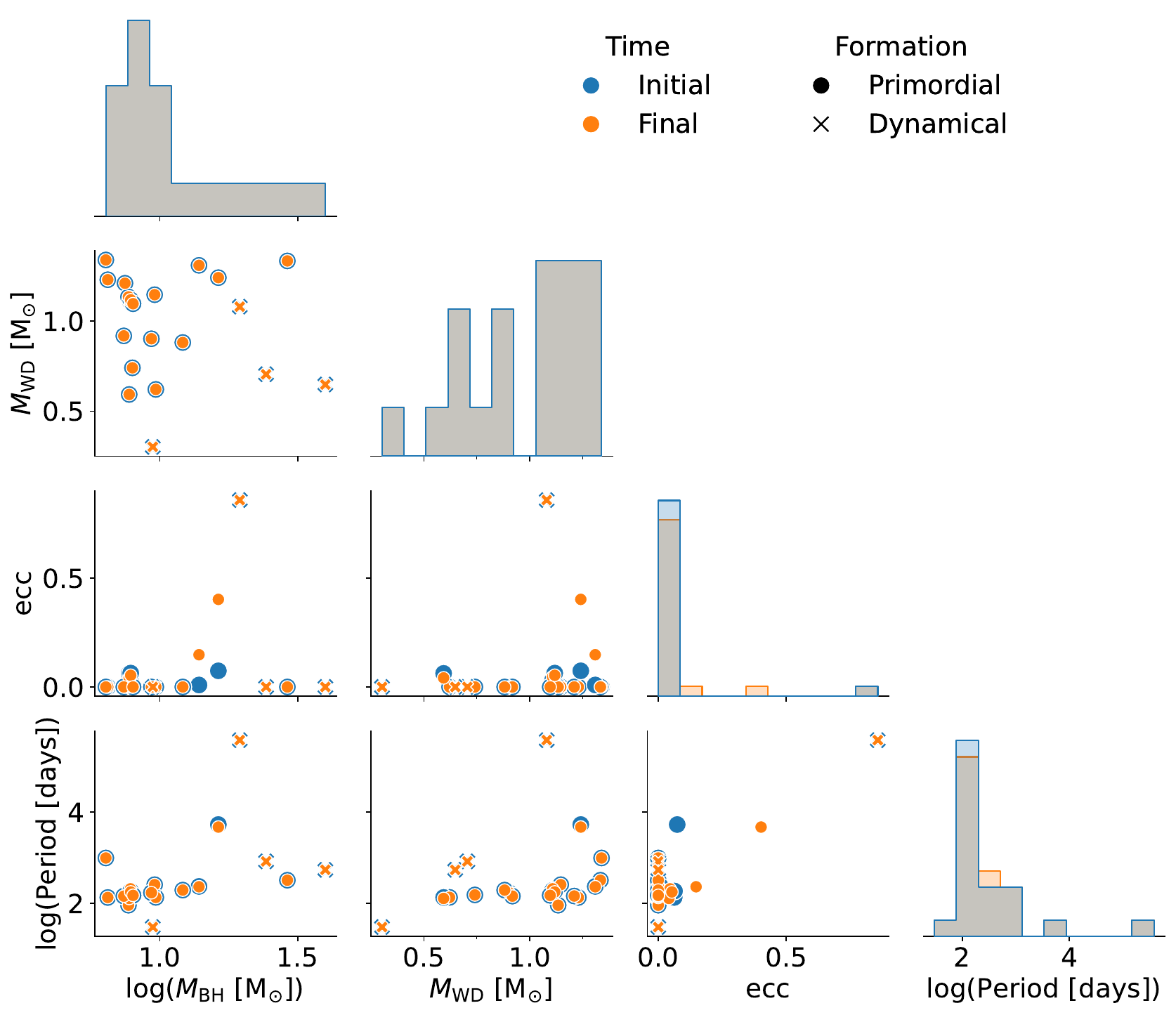}
\caption{Properties of the BH-WD ejected binaries, the dots and the crosses represent primordial binaries and dynamical binaries respectively; the binaries at the first BH-WD time are shown in blue, the last BH-WD time is represented by orange colour.}
\label{f: BH-WD corner}
\end{figure*}

From  Fig.~\ref{f: BH-GS corner} it is evident that almost all the BH-GS ejected systems are from the primordial binary population with the \gls{bh} mass peak at  $M_{\rm BH} \simeq 8~\rm M_{\sun}$, similarly to the BH-MS case; the eccentricity distribution has a strong peak near $e \simeq 0$. The only systems that  have $e > 0$ have a \gls{gs} mass in the range $M_{*} \in [10,35]~\rm M_{\sun}$. Roughly half of these systems maintain a finite eccentricity up to the end of the BH-GS phase, in the other half the eccentricity drops to $e = 0$. As previously discussed for BH-MS binaries, the evolution towards  $e \simeq 0$ is driven  by tides and \gls{ce} evolution. We find $63\%$ of BH-GS progenitors undergo a \gls{rlo} event that reduces the median of the eccentricities from $ \simeq 0.21$ pre-\gls{rlo} to $\simeq 0$ post-\gls{rlo}, and $37\%$ of the systems undergo a \gls{ce} phase. 

After ejection, $26\%$ of BH-GS undergo a \gls{ce} phase when they are classified as  BH-GS and $49\%$ experience a \gls{rlo} phase. Moreover, in the vast majority of cases, the \gls{rlo} phase occurs when the stellar companion is a \gls{hg} star. 
Unlike the BH-MS case, the number of dynamically formed systems (2) is significantly lower compared to the primordial binaries (33), making it difficult to provide a meaningful comparison between the two populations.

As for the BH-MS binaries, we look for triple systems in our BH-GS  sample. We consider the 35 ejected BH-GS binaries that are presented in Fig.~\ref{f: BH-GS corner} and we find that $37\%$ of the BH-GS binaries have formed a stable triple system during their in-cluster evolution. 
These  systems have an high average outer orbit eccentricity ($e = 0.94$), and an average semi-major axis ratio ($a_{\rm in}/a_{\rm out} = 0.004$).

Among the ejected population, we find  that 13 BH-GS progenitors were in a stable triple system at the moment of ejection.  These systems are the same triple systems with an inner BH-MS inner binary discussed  in Section~\ref{sec:BH-MS}.

The BH-GS retained binaries are shown in  Fig.~\ref{f: BH-GS corner_RET}. The distribution of \gls{bh} masses show the same cut-off at $45~\rm M_{\sun}$ found in the BH-MS case (Fig.~\ref{f: BH-MS corner_RET}). The mass distribution of \glspl{gs} is similar to the one found for BH-MS binaries, although we notice that the low mass part of the distribution is now mostly populated by primordial binaries. 
We note that the three \glspl{bh} that have a mass above the \gls{ppi} limit in the whole sample are part of the in-cluster population 
The most massive among them has a mass of $225~\rm M_\odot$.

\section{BH-WD binaries}
\label{sec:BH-WD}
In this section, we consider ejected BH-WD binaries. \gls{wd} stars are defined following the $\tt BSE$ classification (indices 10, 11 and 12). The results are reported  in Fig.~\ref{f: BH-WD corner}.

The number of  BH-WD binaries formed is given in Table \ref{tab:models}; $22\%$ of the ejected BH-WD binaries are formed before the ejection, the remaining are formed during the isolated evolution after ejection. Fig.~\ref{f: BH-WD corner} shows that, as expected, the properties of the binaries remain essentially the same during the lifetime of the systems. The \gls{bh} mass distribution is peaked around $M_{\rm BH} \simeq 8 ~\rm M_{\sun}$, with the exceptions being massive, dynamically formed systems. The \gls{wd} masses are concentrated around $M_{\rm WD} \simeq 1.1~\rm M_{\sun}$, with an extended lower mass tail. However, we note that this peak is likely strongly dependent on the max simulated time of our clusters ($3~\rm Gyr$) since there are many more low mass stars which have not had time to collapse to a \gls{wd}. Similarly to the BH-MS and BH-GS systems, the eccentricity has a strong peak at $e \simeq 0$, and, as before, the reason are \gls{ce} phase and tidal friction: $68 \%$ of the BH-WD progenitors experience at least one \gls{rlo} event and  $67\%$ of the systems undergo a \gls{ce} event before ejection, which reduces the eccentricity from a median $ \simeq 0.18$ to $e \simeq 0$. During almost all the \gls{rlo} events, the \gls{ce} phase is present.

The percentage of BH-WD progenitors that undergo at least one \gls{ce} is higher than for BH-MS and BH-GS. BH-WD binaries are more likely to undergo  several \gls{ce} and \gls{rlo} phases during their evolution, therefore their eccentricity distribution is more strongly peaked at  $e\sim 0$ than BH-MS and BH-GS binaries. Although, we note that the previously mentioned caveat should be considered here since BH-WD binaries that are formed after the current max simulation time will most likely form dynamically between a lone \gls{wd} and \gls{bh}. These binaries will then not experience \gls{ce} or \gls{rlo} phases.

The \gls{ce} phase and tides alone could fully explain the eccentricity and period distributions in Fig.~\ref{f: BH-WD corner}. 
There are two exceptions: a dynamically formed system and a primordial one. The former is formed dynamically and its separation is sufficiently high to avoid \gls{ce} or strong tides, the latter is formed inside the stellar cluster and it is discussed below. We find that $19\%$ of BH-WD systems are formed dynamically. 

We study the possible impact of a tertiary in the formation of the ejected population of BH-WD systems: 4 ($19 \%$) BH-WD binaries have been in a stable triple system during their in-cluster evolution, with all the inner binaries from the primordial binary population. Only 1 BH-WD system progenitor was found in a stable triple system at the moment of ejection. The  semi-major axis ratio at the last time the triple existed in the cluster model  is $a_{\rm in}/a_{\rm out} = 0.04$, and the  eccentricity of the outer orbit is $e = 0.95$.

 The highly eccentric primordial binary mentioned previously has a stable triple during its in-cluster evolution: this system avoids a \gls{ce} phase and the orbit becomes highly eccentric due to triple dynamics. In the other 3 stable triples, the inner binary undergoes a \gls{ce} phase during which its orbit circularises and the period decreases. Because the distance of the tertiary to the inner binary is relatively large, it seems unlikely that after a \gls{ce} phase the presence of the tertiary companion can significantly affect the binary evolution. 

It should be stressed that the importance of the dynamical effects is strongly dependent on the characteristics of the triple system, such as the mass of the tertiary, the separation and the inclination. 
Moreover, the relativistic precession of the inner binary orbit can suppress the eccentricity evolution \citep{2002ApJ...578..775B}, but this effect is not included in our simulations. 

Finally, we looked for retained BH-WD binaries, and only found one dynamical binary with $M_{\rm WD}=0.65~\rm M_{\sun}$, $M_{\rm BH}=67.92~\rm M_{\sun}$, $e=0.5$ and ${\rm period} = 10~\rm days$. Most BH-WD binaries that are formed in the models are immediately disrupted or ejected by dynamical encounters.

\begin{figure*}
\includegraphics[width=1.6\columnwidth]{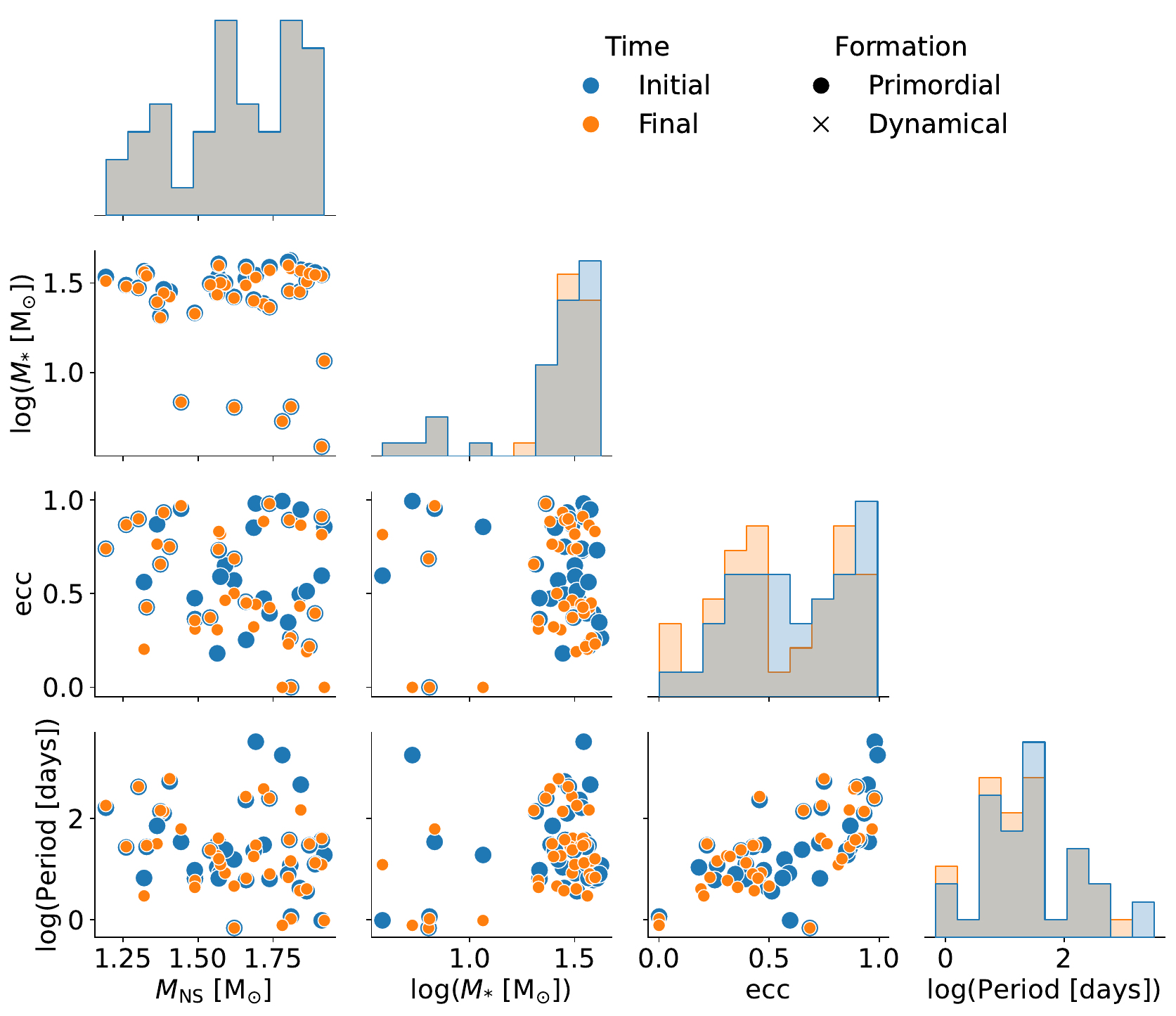}
\caption{Properties of the ejected NS-S  binaries. The binaries at the  time the system is first classified as a NS-S  are shown in blue, the last NS-S time is shown in  orange. Note the absence of dynamically formed binaries.}
\label{f: NS-S corner}
\end{figure*}

\begin{figure*}
\includegraphics[width=2\columnwidth]{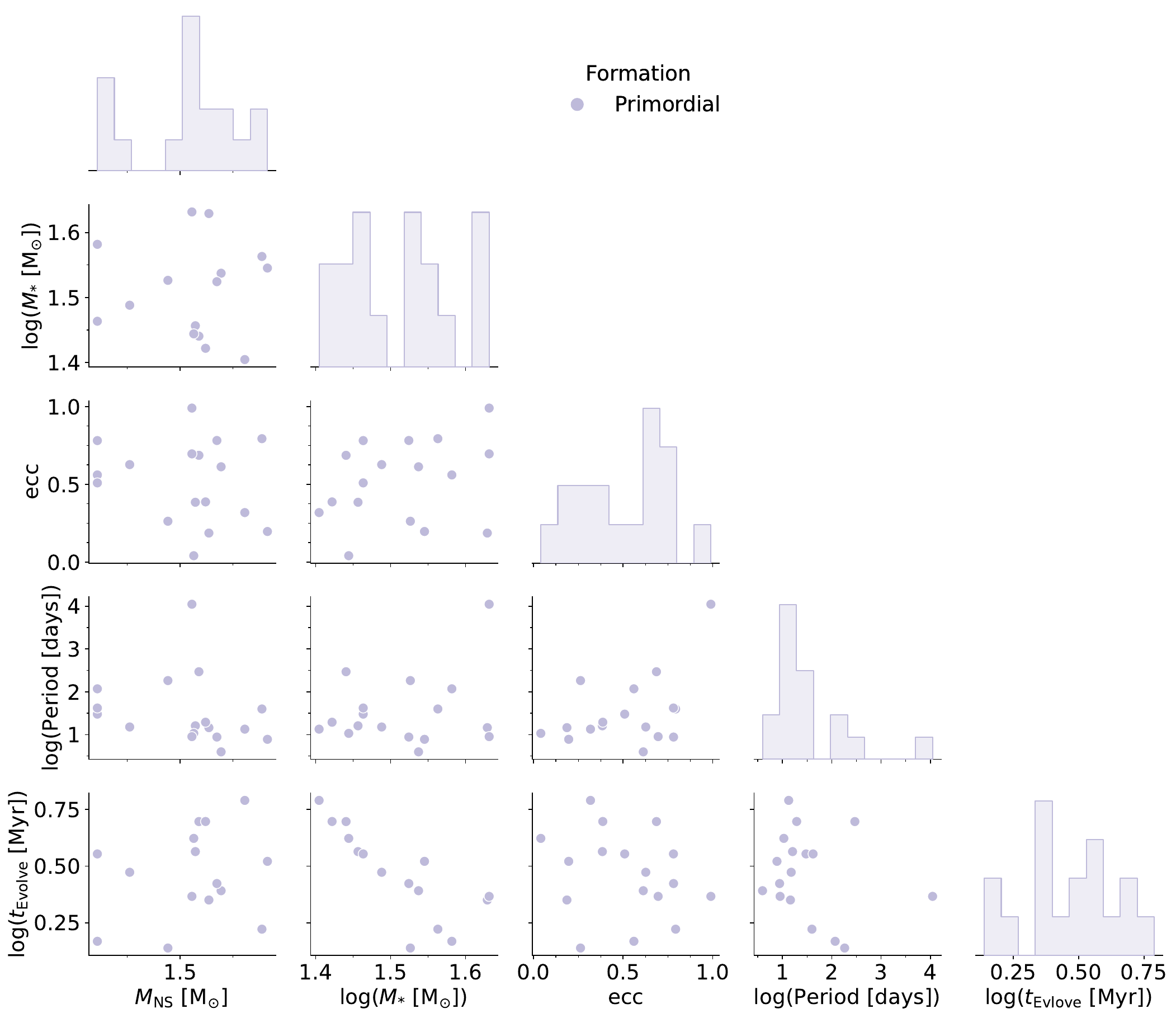}
\caption{Properties of the retained NS-S  binaries. We do not show the dynamically formed binaries as there are none in the sample. Here $t_{\rm evolve}$ represents the lifetime of the binary, i.e., the time between formation and 
 when the stellar component evolves off the MS or giant phase.}
\label{f: NS-S corner_RET}
\end{figure*}

\section{NS-S Binaries}
\label{sec:NS-S}
In this section we consider NS-S binaries, {where we account for both \gls{ms} and \gls{gs} companions.} {The binary properties for the ejected population are shown in Fig.~\ref{f: NS-S corner}}.
We find that all but two of the ejected NS-S systems are formed before the ejection from the cluster, { and that they all come from the primordial binary population.}  The latter point can be explained by considering that it is more difficult for a \gls{ns} to capture a companion through dynamical interactions as the central region of the cluster, where these interactions typically occur, is dominated by \glspl{bh} until the end of the simulation. This difficulty in forming NS-S binaries dynamically in star clusters has also been found by previous work~\citep{Tanikawa2024CompactBF}. 

The \gls{ns} masses are distributed uniformly across the interval $M_{\rm NS} \in [1, 2]~\rm M_{\sun}$, while the stellar mass distribution of the companion is peaked around $M_{*} \simeq 30~\rm M_{\sun}$, with some less massive stars down to $M_{*}\approx3~\rm M_{\sun}$. The eccentricity distribution is closer to uniform compared to the BH-GS or BH-WD cases (see Section~\ref{sec:BH-GS} and Section~\ref{sec:BH-WD}). Several factors must be accounted for when considering the eccentricity distribution. 

While NS-S progenitors maintain an almost circular orbit throughout  their evolution, the \gls{ns} natal kick increases the eccentricity of the binaries when the \gls{ns} is formed. For this reason, we expect higher eccentricities than  for BH-GS and BH-WD binaries. In the latter systems, \gls{ce} and \gls{rlo} occur shortly before the {\gls{bh} forms which circularises the binary}, and then the reduced \gls{bh} natal kick is not sufficient to produce a large spread in the eccentricity distribution. Moreover, due to the fallback prescription, a significant fraction of \glspl{bh} can be formed without a natal kick.   

We find that $37\%$ of the ejected NS-S systems undergo a \gls{ce} phase and that $97\%$ experience a \gls{rlo} event. In essentially all cases, the \gls{rlo} and \gls{ce} phases occur when the companion star is on the \gls{hg}, and often the binary does not survive up to the \gls{gs} phase of the companion. These behaviours explain the absence of the peak at $e\simeq 0$ in Fig.~\ref{f: NS-S corner}.

{Although we find no ejected NS-S binaries in stable triple systems at the time of ejection, we do find that the $32\%$ of the progenitor systems have been part of a stable triple system inside the cluster.}
We investigate these progenitor triple systems and find that they have an average semi-major axis ratio of $a_{\rm in}/a_{\rm out} = 0.0001$ -- the smallest through all the type of triple systems studied in this article. In fact the average semi-major axis of the inner binary ($ \simeq 100~\rm R_{\sun}$) is more than five times smaller than for all the other triple systems previously mentioned ($ \simeq 580~\rm R_{\sun}$). Moreover, these NS-S progenitor triples have the largest average outer semi-major axis across all other triple systems, and the average outer eccentricity is $e_{\rm out} = 0.96$.

We show the properties of the retained NS-S binaries in Fig.~\ref{f: NS-S corner_RET} {and similar to the ejected population, we find only systems from the primordial binary population}. The distribution of the \gls{ns} masses is uniform between $1.1~\rm M_{\sun}$ and $1.9~\rm M_{\sun}$, while the distribution of stellar masses (both \gls{ms} and \gls{gs}) is peaked around $30~\rm M_{\sun}$. The binary eccentricities are spread between 0 and 1, with a slight preference for high eccentricities likely due to the \gls{ns} natal kick. {The binary periods are found predominantly between $1~\rm day$ and $100~\rm days$, peaking at around $\approx18~\rm days$. We only find 4 systems exceeding $100~\rm days$. 

\section{Application to Gaia Black holes}
\label{sec:App_GaiaBH}
In this section we investigate the formation of Gaia \gls{bh}-like  systems \citep{El-badry2022,El-Badry2023,El_Badry_2024} in our simulations: {We identify the top candidates within the sample, outline the formation pathways of these systems, and analyse the formation efficiencies of Gaia-like systems in our models.}  Similar studies can be found in \cite{Mar_n_Pina_2024}, \cite{rastello2023dynamicalformationgaiabh1}, \cite{Di_Carlo_2024} and \cite{Tanikawa_CBCForm}.

To quantify the capability of forming a certain type of systems in our simulations, we  define the formation efficiency as:
\begin{equation}
    \eta = N_{\rm BH-type} / {M_{\rm tot}} ~ ,
    \label{Efficiency}
\end{equation}
where $N_{\rm BH-type}$ is the number of binaries of a certain category (BH-MS, BH-GS or Gaia-like binaries) and $M_{\rm tot}$ is the total mass of the cluster in our simulations or of a certain type (e.g., clusters with the same metallicity or density).
Here, Gaia-like binaries mean all  binaries that fall within the contour similarity regions given in Fig.~\ref{f: BH-MS corner} and Fig.~\ref{f: BH-GS corner}. As shown in these figures, there is no ejected binary that satisfies all observational constraints at once. It is also evident  that the constraint on the eccentricity poses a strong limit for our sample. If we relax this analysis by removing completely the eccentricity constraint, we find that there are two systems that are within the Gaia BH1 similarity region, and two systems that lie within the Gaia BH2 similarity region.  However, even after relaxing the eccentricity constraint, we do not find  any Gaia BH3-like system and therefore only consider Gaia BH1 and BH2 in what follows.

The relaxation of the eccentricity constraint is justified by the considerable uncertainty in the prescription for binary eccentricity evolution within the $\tt BSE$ framework.
Finally we stress that for each system, we consider it within the similarity region if at any point during the evolutionary stage it is within the contours and not necessary at the first/last moment. 

The two Gaia BH1-like ejected systems both have a dynamical formation.
The first system is a dynamically formed binary from a dense ($\rho_{\rm h} = 10^5~\rm \density$), relatively low mass ($M_{\rm c} = 10^{4}~\rm M_{\sun}$) cluster, with a sub-solar metallicity ($Z = 0.001$) and an initial binary population. The second system is also formed dynamically, and it is ejected from a low density ($\rho_{\rm h} = 1200~\rm \density$), intermediate mass ($M_{\rm c} = 10^{5}~\rm M_{\sun}$) cluster, with sub-solar metallicity ($Z = 0.001$) and zero initial binary fraction (i.e. no primordial binaries). The second system is also the `closest' to Gaia BH1 in parameter space and it is further considered in \ref{sec:Gaia_BH1 closest}. 
The efficiency for Gaia BH1-like systems considering the entire mass of the clusters ($M_{\rm tot} \simeq 4.6 \times 10^6~\rm M_{\sun}$) is $\eta  = 4.36 \times 10^{-7}~\rm M_{\sun}^{-1}$ which is comparable to the value found by \cite{rastello2023dynamicalformationgaiabh1} in clusters with lower mass and density.

The two Gaia BH2-like systems have a primordial origin: the first system formed in a dense, intermediate mass cluster ($\rho_{\rm h} = 10^5~\rm \density$, $M_{\rm c} = 5 \times 10^4~\rm M_{\sun}$), with sub-solar metallicity ($Z = 0.001$). This system is described in detail in section \ref{sec:closest_GaiaBH2}. The second system is also a primordial binary ejected from an intermediate dense cluster ($\rho_{\rm h} = 10^4~\rm \density$), with mass $M_{\rm c} = 10^5~\rm M_{\sun}$, sub-solar metallicity ($Z = 0.001$). Considering the absence of a proper eccentricity constraint for these systems, it is important to underline that both  Gaia BH2-like systems undergo at least one \gls{rlo} and \gls{ce} phase during their evolution. In particular, the second system escapes the cluster  due to a \gls{bh} natal kick. Its eccentricity at ejection is  $e \simeq 0.5$,  and it is subsequently circularised during a \gls{rlo} event. 
We find that the formation efficiency for Gaia BH2-like systems is also $\eta = 4.36 \times 10^{-7}~\rm M_{\sun}^{-1}$.

We quantify the formation efficiency of our models in producing BH-S binaries in Fig.~\ref{Eff_new}. There we show the total production efficiency for all (ejected and retained) BH-S binaries {and also for only ejected BH-S binaries. We then further plot the formation efficiency of ejected BH-MS and BH-GS binaries separately.} We plot these efficiencies both as a function of $Z$ and $M_{\rm c}$, and separating clusters by $\rho_{\rm h}$. The efficiency shows a strong peak around $Z = 10^{-3}$, and decreases in high mass clusters for both ejected and retained binaries. There are no ejected or retained BH-S binaries in the highest mass clusters and there are two main reasons for this. First, the high mass of the cluster means a higher escape velocity, making it more difficult for binaries to be ejected in dynamical interactions. Secondly, the most massive simulations ($M_{\rm c} \geq 5 \times 10^5~\rm M_{\sun}$) ran for a shorter time than the less massive ones ($ \leq 608~\rm Myr$), which leads to a lower number of binaries being produced dynamically per unit mass.
Furthermore, we note that the efficiency production of ejected BH-S binaries increases with cluster density, but it remains nearly independent of density for the retained population. This suggests that retained BH-S binaries mostly form from the evolution of the primordial binary population with little effect from dynamics. {Our previous analysis of the retained BH-MS and BH-GS binaries (Fig.~\ref{f: BH-MS corner_RET} and Fig.~\ref{f: BH-GS corner_RET} respectively) supports this since we show that they are predominantly formed from the primordial binary population.}

\begin{figure*}
    \centering
    \includegraphics[width=0.9\textwidth]{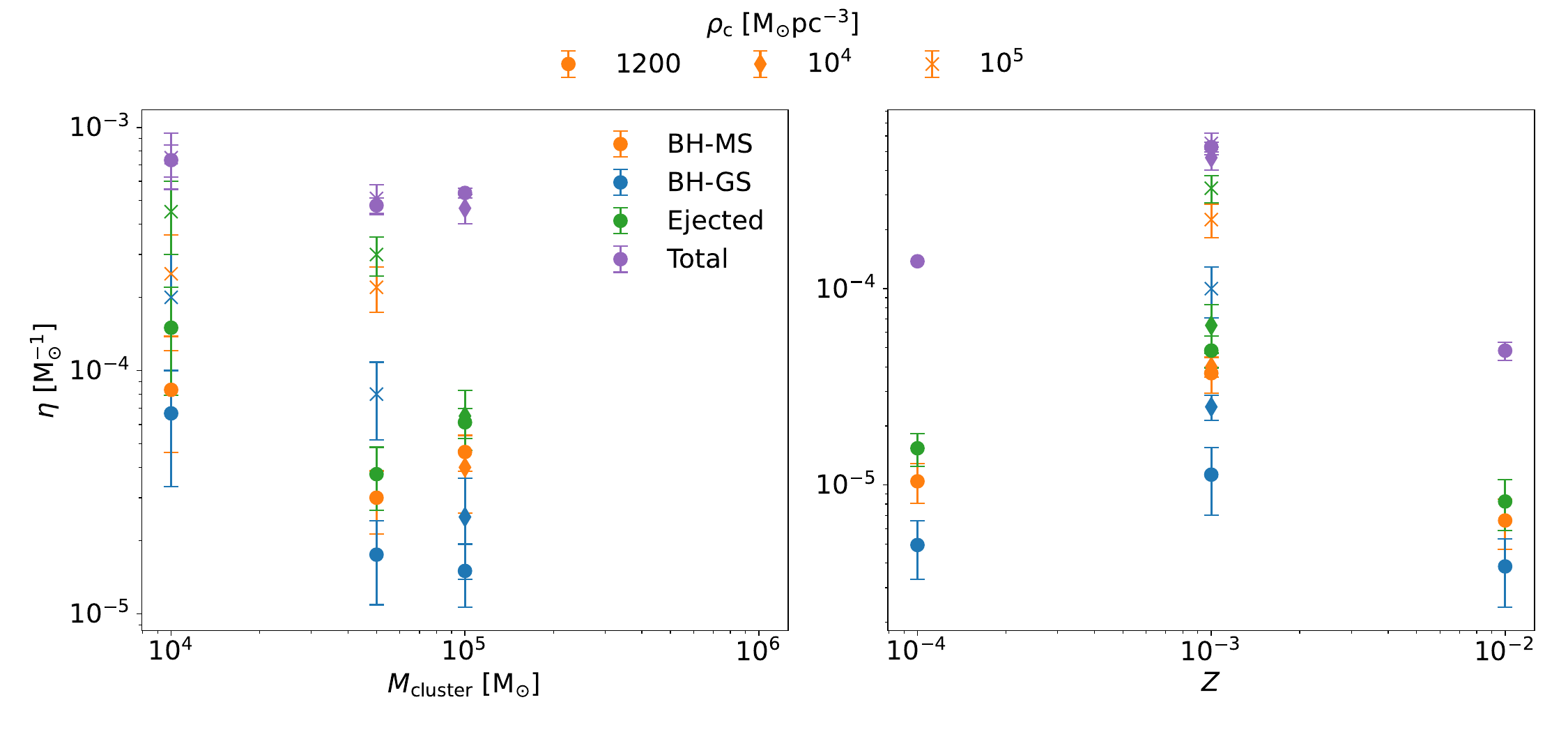}
        \caption{Here we show the formation efficiency of the ejected BH-S binaries split into BH-MS and BH-GS systems, as a function of initial cluster mass (left panel) and cluster metallicity (right panel). We distinguish between initial cluster half-mass density by varying marker symbols. For each cluster type we also show the formation efficiency for the ejected binary population and the total population (ejected + retained binaries).    
        }
    \label{Eff_new}
\end{figure*}

\subsection{Best Gaia BH1 match}
\label{sec:Gaia_BH1 closest}
To identify the binary in our model that most closely resembles Gaia BH1, we follow \cite{Di_Carlo_2024}. The distance between a point in the parameter space and a Gaia system is defined as $ |x - x_{\rm Gaia}| / x_{\rm Gaia}$, where $x$ is the array of parameters of the binary and $x_{\rm Gaia}$ are the parameters for Gaia BH1 reported in Table \ref{tab:Resume_Gaia}. Following  \cite{Di_Carlo_2024}, we do not take the eccentricity of the binary into account for distance computation, as we assume that the eccentricities remain uncertain during $\tt BSE$ evolution as stated earlier.

The parameters of the closest system to Gaia BH1 are reported in  Table \ref{tab:Closest_Gaia}. The system is a dynamically formed BH-MS binary which is formed early in the cluster ($\sim 6~\rm Myr$) due to the encounter between a $12.7~\rm M_{\sun}$ naked He star and a low mass \gls{ms} star ($M_{*} \simeq 0.5~\rm M_{\sun}$). The He star rapidly ($t \simeq 0.02~\rm Myr$) explodes in a supernova, producing a kick that expands the orbit (from $a \simeq 32~\rm R_{\sun}$ to $a \simeq 175~\rm R_{\sun}$) and makes it highly eccentric (from $e \simeq 0$ to $e \simeq 0.85$). The \gls{sn} remnant is a stellar-mass \gls{bh} with $M_{\rm BH} \simeq 9.5~\rm M_{\sun}$. The system exists in the cluster for only $\approx1~\rm Myr$ before it is ejected, with velocity $v_{\rm esc} = 125.3~\kms$, and it continues to evolve in isolation. The binary then survives for nearly a Hubble time with its orbital properties largely unchanged. The evolution is schematically illustrated in Fig.~\ref{fig:BH1_path}. The estimate on the age of Gaia BH1 form \cite{Tanikawa_CBCForm} is $ \gtrsim 1~\rm Gyr$. This is also  compatible with our best candidate, whose properties after ejection are essentially unchanged up to a Hubble time.

\begin{figure} 
    \centering
        \includegraphics[width=0.8\linewidth]{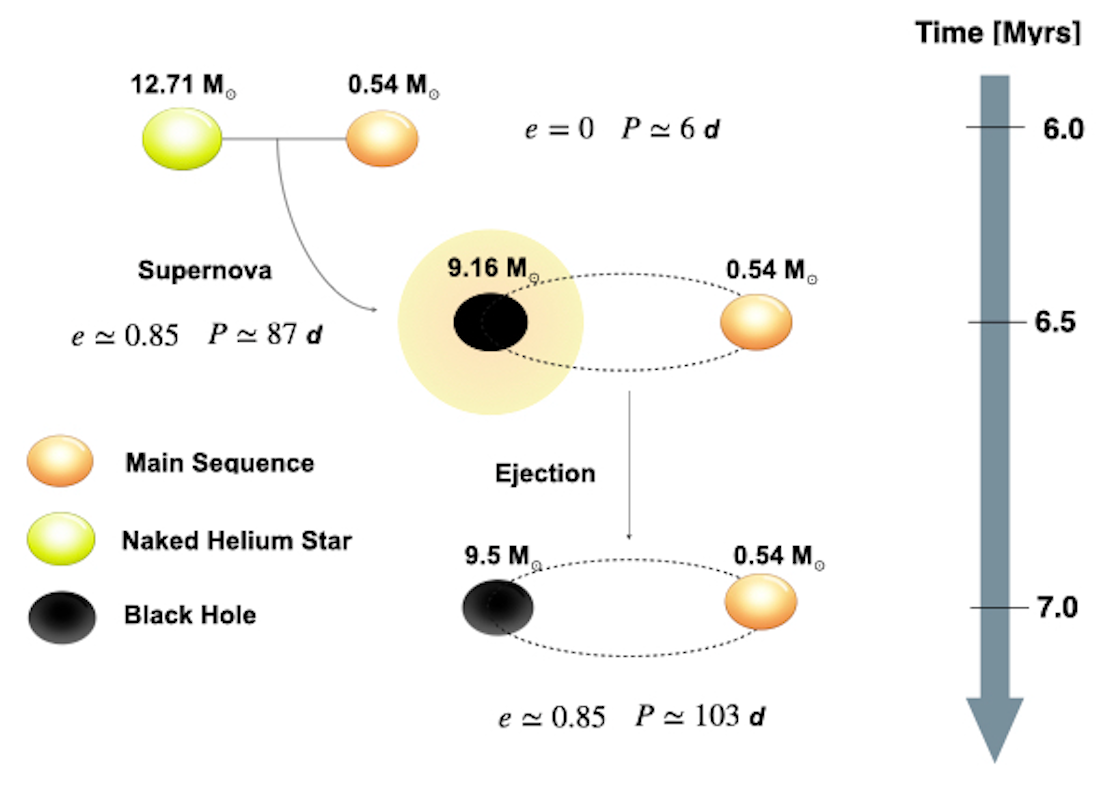}
        \caption{Formation pathway of the closest Gaia BH1-like system which was found in model Z2-M10-D3.}
        \label{fig:BH1_path}
\end{figure}

\begin{table*}
\centering
\begin{tabular}{ccccccccc} \hline
       & $M_{\rm BH}~\rm[M_{\sun}]$  & $M_{*}~\rm[M_{\sun}]$ &  $a~\rm[R_{\sun}]$ &  $P~\rm[days]$ & $e$ & $M_{\rm c}~\rm[M_{\sun}]$ & $\rho_{\rm c}~\rm[\density]$ & $Z$ \\
\midrule
Closest to Gaia BH1  & $9.16$  & $0.54$ & $197.32$ & $103.12$ & $0.85$ & $10^5$ & $1200$ & $0.001$\\[0.1cm]
Closest to Gaia BH2 &  $9.27$ & $1.07$ & $208.28$ & $108.31$ & $0.0$ & $5 \times 10^4$ & $10^5$ & $0.001$\\[0.1cm]
Closest to Gaia BH3 & $9.55$  & $0.76$ & $187.86$ & $93.01$ & $0.0$ & $10^5$ & $1200$ & $0.01$ \\
\bottomrule
\end{tabular}
\caption{\label{tab:Closest_Gaia} Properties of the closest to Gaia \gls{bh} systems as well as the key properties of the cluster where they are formed.}
\end{table*}

\begin{figure} 
    \centering
        \includegraphics[width=0.8\linewidth]{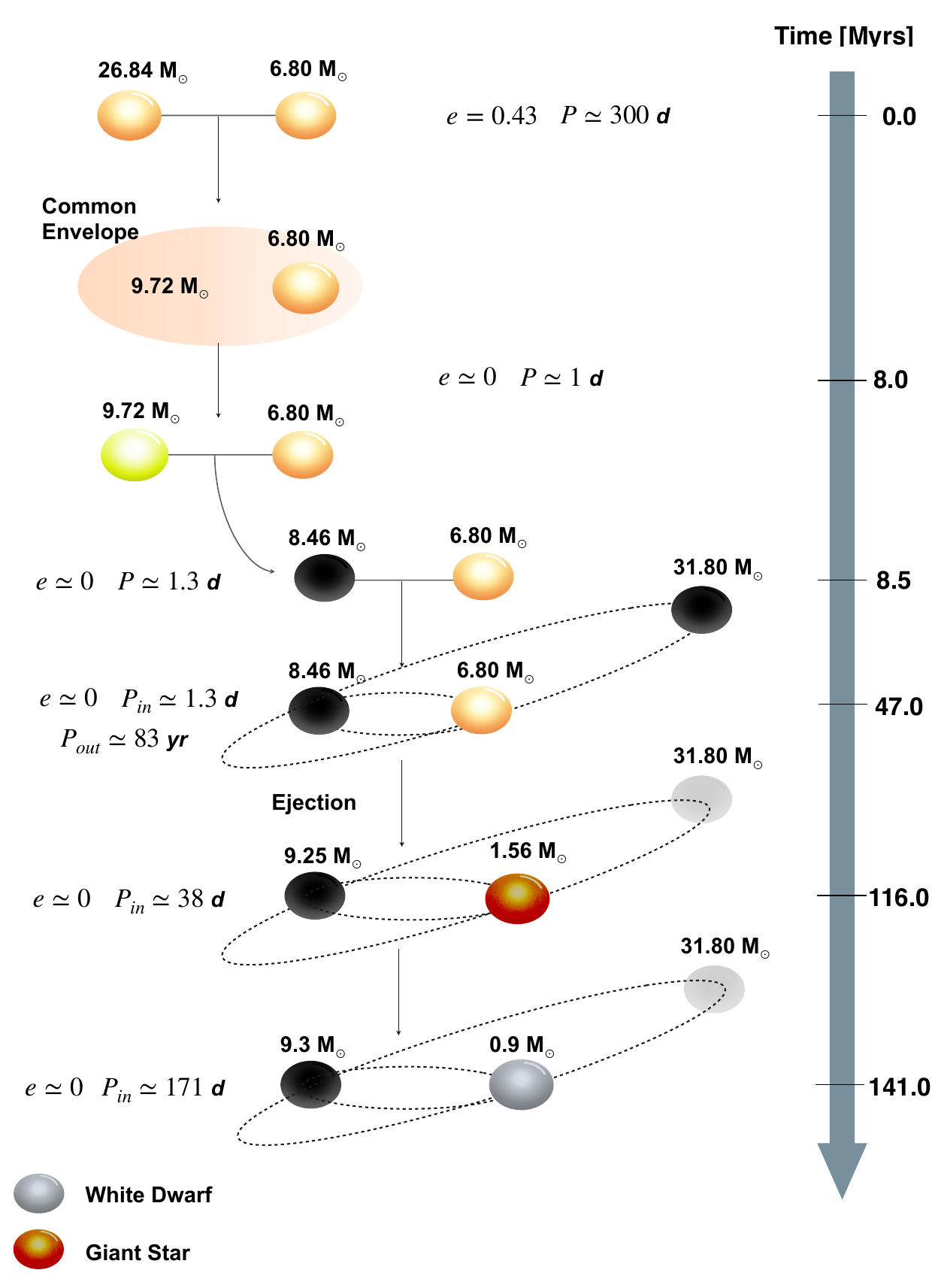}
        \caption{Formation pathway of the nearest Gaia BH2-like system; the inclination of the orbit is represented symbolically, with the actual inclination being $25^{\circ}$. The tertiary component after ejection is shaded to emphasise that evolution continues only for the inner binary, while the tertiary is disregarded post-ejection. This system is found in model Z2-M5-D5.}
        \label{fig:BH2_path}
\end{figure}

\subsection{Best Gaia BH2 match}
\label{sec:closest_GaiaBH2}
In this section, we describe the closest system to Gaia BH2, with the same criteria as in Section~\ref{sec:Gaia_BH1 closest}. The parameters of the closest system to Gaia BH2 are reported in Table~\ref{tab:Closest_Gaia}.

The system is formed as a primordial binary in a dense cluster with $\rho_{\rm h} = 10^5~\rm  \density$, $M_{\rm c} = 5 \times 10^4~\rm M_{\sun}$, and $Z = 0.001$; the orbit has a medium eccentricity ($e = 0.43$) and a semi-major axis $a = 607.5~\rm R_{\sun}$; the masses of the \gls{ms} stars are $M_{1} = 26.84~\rm M_{\sun}$ and $M_{2} = 6.80~\rm M_{\sun}$. The most massive star quickly leaves the \gls{ms}, starts burning Helium in the core evolving into a naked helium star and losing $64\%$ of its mass within $8~\rm Myr$. During this time, the evolving star fills its Roche lobe and then enters a \gls{ce} phase that circularises and shrinks significantly the orbit ($e = 0.0$, $a = 11.2~\rm R_{\sun}$). After $0.5~\rm Myr$, the massive star collapses and forms a \gls{bh} with $M_{\rm BH} = 8.46~\rm M_{\sun}$. Following another $\simeq 38.5~\rm Myr$, the binary forms a stable triple with a tertiary \gls{bh} ($M_{\rm BH} = 31.8~\rm M_{\sun}$). The triple system is wide ($a_{\rm in}/a_{\rm out} \simeq 0.001$) and thus the inner binary orbital properties are not significantly perturbed by the tertiary. The binary is ejected from the cluster (with the tertiary still bound) as a BH-MS after $50~\rm Myr$ with an escape velocity $v_{\rm esc}=11.4~\kms$. Soon after the ejection, the \gls{ms} star crosses the \gls{hg}, moving into the red giant phase and slowly stripping its envelope. During this phase, the winds from the evolving star widen the orbit. Finally, $141~\rm Myr$ from the start of the simulation, the star collapses into a Carbon-Oxygen \gls{wd} with $M_{\rm WD} = 0.9~\rm M_{\sun}$ in a circular orbit ($e \simeq 0$, $a = 281.4~\rm R_{\sun}$, $P = 171.12~\rm days$) with a \gls{bh} of $M_{\rm BH} = 9.3~\rm M_{\sun}$.

The schematic representation of this evolution is shown in Fig.~\ref{fig:BH2_path}. During the red giant phase, the properties of the system (reported at the closest moment in Table~\ref{tab:Closest_Gaia}) are quite different than the ones measured for Gaia BH2. Moreover, the age of the system is $\simeq 116~\rm Myr$, which is also different from the estimated age of Gaia BH2 ($\gtrsim 5~\rm Gyr$).

\section{Discussion}
\label{sec: Discussion} 

We investigated the population of BH-star binaries formed in 32 of $N$-body simulations of dense and massive star clusters.
Compared to previous $N$-body simulations which focused on low mass clusters ($M_{\rm c} \in [10^2, 10^4]~\rm M_{\sun}$), {we have explored more massive clusters in the range $M_{\rm c} \in [10^{4}, 10^{6}]~\rm M_{\sun}$.}
The formation of Gaia BH1 in low mass clusters was investigated in \cite{rastello2023dynamicalformationgaiabh1}. They provide $3.5 \times 10^4$ direct $N$-body simulations of clusters with an initial mass between $3 \times 10^2~\rm M_{\sun}$ and $3 \times 10^4~\rm M_{\sun}$ at solar metallicity ($Z \simeq 0.02)$ and define Gaia BH1 binaries with the similarity regions reported in Section~\ref{sec:BH-MS}. They find one ejected Gaia BH1 like system and compute a formation efficiency from all their models of $\eta \simeq 2 \times 10^{-7}~\rm M_{\sun}^{-1}$. This result is compatible with the value of $\eta \simeq 4 \times 10^{-7}~\rm M_{\sun}^{-1}$ found in our study. However, we note that we have not considered any eccentricity constraint and both our Gaia BH1 candidates come from sub-solar metallicity clusters ($Z = 0.001$). 

\cite{Tanikawa_CBCForm} studied the formation of Gaia BH1-like systems using 100 $N$-body simulations with cluster masses $ \simeq 10^3~\rm M_{\sun}$ and metallicity $Z = 0.005$. They define Gaia BH1-like binaries as those with properties such that $M_{*} \leq 1.1~\rm M_{\sun}$, $P \in [100, 2000]~\rm days$ and $e \in [0.3, 0.9]$ and they find $\eta \simeq 10^{-5}~\rm M_{\sun}^{-1}$. Adopting the same contours (and as before setting no constraint on eccentricity), for our BH-MS sample, we obtain a smaller $\eta \simeq 1.52 \times 10^{-6}~\rm M_{\sun}^{-1}$. As previously underlined in \cite{rastello2023dynamicalformationgaiabh1}, the difference in the initial conditions in \cite{Tanikawa_CBCForm} can have an impact on the number of Gaia \gls{bh}-like systems and efficiencies. In particular, they set a binary fraction of $100\%$ whilst we opt for a total initial binary fraction of $0.25\%$, which is $100\%$ among \gls{bh} progenitors. Both these factors could explain the larger number of Gaia BH1-like systems formed in \cite{Tanikawa_CBCForm}.

Our simulations explore masses and densities  similar to the models in \cite{Mar_n_Pina_2024}, who used the Monte Carlo simulations from \cite{kremer_modeling_2020}. Our total efficiency for high mass clusters ($M_{\rm c} =  10^{5}~\rm M_{\sun}$) is $5\times 10^{-5} ~\rm M_{\sun}^{-1}$, consistent although somewhat smaller than the efficiency of $\sim 10^{-4} ~\rm M_{\sun}^{-1}$  found by \cite{Mar_n_Pina_2024} for similar cluster masses and densities.

Moreover, it is important to consider the impact of different initialisations of the stellar mass distribution across all of these studies. Half of our cluster models are initialised with all stars with $M_{*} \geq 20~\rm M_{\sun}$ in a binary system, whilst in \cite{rastello2023dynamicalformationgaiabh1} all stars with $M_{*} \geq 5~\rm M_{\sun}$ is in a binary. This choice causes the number of low mass binaries with no \gls{bh} progenitor star to be higher in \cite{rastello2023dynamicalformationgaiabh1} compared to our simulations, possibly leading to a higher production efficiency of BH-S binaries through dynamical exchanges. In addition, \cite{rastello2023dynamicalformationgaiabh1} integrated their models for a larger number of initial cluster relaxation time, which was achievable due to their lower initial cluster masses. We note that for many of our models there still exists a significant BH population ($>20$) within the cluster. Thus, these clusters can still be considered "alive" for the purposes of forming new binaries. Continuing the simulation of these clusters would likely produce many more Gaia-like BH systems (especially in our massive clusters with $\mathcal{O}(10^{2})$ BHs remaining in cluster), increasing our formation efficiency.

\subsection{Conclusions}
The primary objective of this study is to characterise the population of ejected binaries from the $\tt PeTar$ cluster simulations presented in \cite{barber2024} (Section~\ref{sec: simulations}). 
We focus our analysis on the ejected populations of BH-MS, BH-GS, BH-WD and NS-S binaries. After the ejection, the binaries are evolved in isolation up to a Hubble time using the $\tt COSMIC$ population synthesis code. 
{The distributions of the component masses and orbital properties for each of the binary types is presented in Figs.~\ref{f: BH-MS corner}, \ref{f: BH-GS corner}, \ref{f: BH-WD corner} and \ref{f: NS-S corner}, where we also make the distinction between the dynamically formed and primordial binaries. In each of these plots we also show the properties at the first and last instance of that type of binary system.} We also provide similar distributions for the retained binaries (Figs.~\ref{f: BH-MS corner_RET}, \ref{f: BH-GS corner_RET} and \ref{f: NS-S corner_RET}). 

We present a qualitative analysis of the effects of \gls{rlo}, \gls{ce} evolution, tidal interactions and the presence of a stable tertiary on the eccentricity distribution of our population. We find that a large fraction of systems (always more than $20\%$) undergo a \gls{ce} phase or a \gls{rlo}  event during the evolution, which lead to orbital circularisation and a peak at low eccentricities in the population  distributions. The presence of a stable tertiary is found to be insufficient to significantly increase the inner binary eccentricity. On the other hand, the \gls{ns} natal kick is the most efficient channel to produce eccentric binary systems with \gls{ns} components.

We look for Gaia \gls{bh}-like binaries in our sample (Section~\ref{sec:App_GaiaBH}). We choose the regions of similarity for the three Gaia \gls{bh} systems following previous literature \citep{rastello2023dynamicalformationgaiabh1,Di_Carlo_2024,Tanikawa_CBCForm,Mar_n_Pina_2024}, and working under the assumption that the eccentricity of our systems is strongly influenced by the approximated prescription for \gls{ce} and tides in $\tt BSE$ thus highly uncertain.

We present plausible candidates for Gaia BH1 and Gaia BH2 in our ejected sample, as well as a schematic their formation pathways shown in Fig.~\ref{fig:BH1_path} and Fig.~\ref{fig:BH2_path}. We find that Gaia BH1-like systems can form dynamically in a dense and massive star cluster. For Gaia BH2-like systems, we find two binaries that are consistent with the properties of Gaia BH2. These binaries are formed from the primordial binary population and are not assembled dynamically. However, we find all Gaia-like binaries are formed in low-metallicity clusters, which is in contrast with the higher metallicity ([Fe/H]$\sim -0.2$) of the actual observed Gaia BH1 and Gaia BH2 systems. We do not find any binary that is consistent with the properties of Gaia BH3, however see \cite{Mar_n_Pina_2024}. 

Our simulations indicate that clusters can generate a diverse population of binaries consisting of \gls{bh} and stellar components. Many of these binaries align with at least some of the properties observed in Gaia \gls{bh} binaries. 
The fact that only a limited number of our simulated systems align with all the characteristics of the observed systems is likely due to low-number statistics, given that our models generated only 120 ejected \gls{bh}-stellar binaries, along with considerable theoretical uncertainties in modelling binary stellar evolution. Nevertheless, two key findings stand out. Firstly, most of the ejected \gls{bh}-stellar binaries in our models originate from the primordial binary population {($88\%$ for BH-MS and $94\%$ BH-GS binaries)}, rather than being formed dynamically. Secondly, dynamically formed binaries tend to occupy extreme regions of the parameter space, with distributions extending toward higher values of orbital periods and \gls{bh} masses compared to primordial binaries. Moreover, the stellar masses in dynamically formed systems include both the heaviest and the lightest stars in our sample.

Finally, we investigate the dependence of the formation efficiency $\eta$ as a function of cluster properties (section \ref{sec: Discussion}) finding comparable overall results with previous work by \cite{rastello2023dynamicalformationgaiabh1} and \cite{Mar_n_Pina_2024}. Furthermore, we found a strong dependence of the formation efficiency on cluster metallicity and mass: a decreasing trend in accordance with \cite{Mar_n_Pina_2024}, but with slightly lower values of $\eta$.

\section*{Acknowledgements}
We thank Long Wang for useful discussions and for assisting with setting up and running the $N$-body simulations.
Simulations in this paper utilised the high-performance $N$-body code $\tt PeTar$ (version 1047-\_290) which is free available at \url{https://github.com/lwang-astro/PeTar}. We acknowledge the support of the Supercomputing Wales project, which is part-funded by the European Regional Development Fund (ERDF) via Welsh Government. JB is supported by the STFC grant ST/T50600X/1.
FA and FD are supported by the UK’s Science and Technology Facilities Council grant ST/V005618/1.

\section*{Data Availability}
The data used for this work will be freely shared upon reasonable request to the author.



\bibliographystyle{mnras}
\bibliography{Refs} 



\bsp	
\label{lastpage}
\end{document}